\documentclass[a4paper,11pt]{article}
\usepackage{ragged2e}
\usepackage{jcappub}
\usepackage[utf8]{inputenc}
\usepackage[normalem]{ulem}
\usepackage{subfigure}
\usepackage{epsfig} 
\usepackage{amsmath} 
\usepackage{url}
\usepackage{hyperref}
\usepackage{amsfonts}  
\usepackage{float}    
\usepackage{amssymb}   
\usepackage{slashed} 
\usepackage{placeins}
\usepackage{mathrsfs}
\usepackage{color}       
\usepackage{pbox}
\usepackage{array,multirow,makecell}
\usepackage{natbib} 
\usepackage{lipsum}
\usepackage{appendix}
\allowdisplaybreaks

\newcommand{\lsim}[1]{\lesssim}

\makeatletter
\gdef\@fpheader{}
\makeatother

\usepackage{orcidlink}
\begin{document}
\title{Freeze-in $SU(2)$ vector dark matter  at low reheating temperature}


\author[a]{Dilip Kumar Ghosh~\orcidlink{0000-0003-1764-1632},}
\emailAdd{tpdkg@iacs.res.in}
\affiliation[a]{School of Physical Sciences, Indian Association for the Cultivation of Science, \\
2A $\&$ 2B, Raja S.C. Mullick Road, Kolkata, West Bengal-700032, India} 

\author[b]{Sourav Gope~\orcidlink{0000-0003-1165-9452},}
\emailAdd{sourav@phy.iitkgp.ac.in}
\affiliation[b]{Department of Physics, Indian Institute of Technology Kharagpur, Kharagpur 721302, India}

\author[c,d]{Xiao-Gang He~\orcidlink{0000-0001-7059-6311},}
\emailAdd{hexg@phys.ntu.edu.tw}
\affiliation[c]{State Key Laboratory of Dark Matter Physics, Tsung-Dao Lee Institute \& School of Physics and Astronomy, Shanghai Jiao Tong University, Shanghai 200240, China}
\affiliation[d]{Key Laboratory for Particle Astrophysics and Cosmology (MOE) \& Shanghai Key Laboratory for Particle Physics and Cosmology, Shanghai Jiao Tong University, Shanghai 200240, China}

\author[c,d]{Xuan Hong,~\orcidlink{0009-0009-8123-6288}}
\emailAdd{hxxh123@sjtu.edu.cn}

\author[c,d]{Sk Jeesun~\orcidlink{0009-0005-2344-9286}}
\emailAdd{jeesun@sjtu.edu.cn}

\abstract{
The freeze-in mechanism for dark matter (DM) requires extremely feeble interactions with the Standard Model (SM), preventing thermal equilibrium in the early Universe and typically evading experimental detection. However, for sufficiently low reheating temperatures ($T_{\rm RH}$), the observed relic abundance can be realized with larger couplings, opening prospects for experimental searches. In this work, we investigate freeze-in production of $SU(2)_{\rm HS}$ vector dark matter (VDM) in a low-$T_{\rm RH}$ cosmology. The framework naturally contains three mass-degenerate stable VDM candidates without the need for any additional discrete symmetry. We perform a systematic study of the dark matter phenomenology and identify the parameter space consistent with the observed relic abundance. We work with the approximation of instantaneous reheating and consider the DM abundance to be negligible at $T_{\rm RH}$. In contrast to conventional freeze-in scenarios, the required DM couplings can be sizable, rendering part of the parameter space already constrained by existing direct searches like PandaX-4T and LZ, while a significant region remains within the reach of future experiments such as DARWIN.  Though one can realize the freeze-in mechanism for an abelian $U(1)_X$ vector DM models as well, we find that the non-abelian structure of the $SU(2)_{\rm HS}$ scenario leads to a distinct feature due to a larger number of dark matter particles, resulting in an enlarged viable parameter space due to the multiplicity of dark matter states.}

\maketitle

\section{Introduction}
Numerous observations at cosmological and astrophysical scales have established the presence of a non-luminous, non-baryonic dark matter (DM) component in the universe, saturating $25\%$ of the total energy density of the universe\cite{Zwicky:1933gu,Rubin:1970zza,Clowe:2006eq,Planck:2018vyg}.
Since the Standard Model (SM) lacks a viable DM candidate, various beyond-the-SM scenarios have been proposed~\cite{Jungman:1995df,Bertone:2004pz,Feng:2010gw}.
A well-studied example is the weakly interacting massive particle (WIMP), where the relic abundance is set by thermal freeze-out with electroweak-scale interactions~\cite{Arcadi:2017kky,Roszkowski:2017nbc}. However, null results from direct, indirect, and collider searches have significantly constrained the conventional WIMP paradigm~\cite{PandaX-II:2017hlx,PandaX:2018wtu,XENON:2020kmp,HESS:2016mib,ATLAS:2017bfj}.
Another widely explored alternative scenario is the  {\it freeze-in} DM where extremely feeble couplings ($\lesssim {\cal O}(10^{-10})$) of DM prevent thermal equilibrium with the SM bath~\cite{Hall:2009bx,Bernal:2017kxu}.
In this scenario, DM is produced non-thermally via particle decays or annihilations, with an initial abundance that is negligible.
Despite the elegance in the model construction, freeze-in DM models are notoriously difficult to test in the ongoing DM search experiments.
Also, the ad hoc assumption of a negligible DM density in the early universe is now being seriously challenged \cite{Cosme:2023xpa}.

It is interesting to note that the production of any species in the early universe is strongly dependent on the evolution of the universe. The impact of a low reheating temperature on DM production was explored in early studies, e.g., Ref.~\cite{Khalil:2008kp}, within the context of sterile neutrino DM. More recently, it has been demonstrated that the requirement of highly suppressed couplings for freeze-in DM can be substantially alleviated in scenarios involving a low reheating temperature ($T_{\rm RH}$) \cite{Cosme:2023xpa}. 
The reheating temperature is defined as the epoch when the inflaton field dilutes most of its energy density into the radiation bath, making it a subdominant component and creating a radiation-dominated universe.
On the other hand, we have the precise information of a radiation-dominated universe only after the epoch of Big Bang Nucleosynthesis(BBN) at 
the temperature $ \sim {\rm few~ MeV}$~\cite{Kawasaki:2000en,Ichikawa:2005vw}.
Hence, in principle, one can assume $T_{\rm RH}$ as low as $\mathcal{O}(10)$ MeV.
In such a scenario, when the reheating temperature is smaller than the DM mass ($m$),  only a sub-fraction of the particles annihilating to DM can potentially generate DM that lies in the tail of the Boltzmann distribution function with energy $\gg T_{\rm RH}$
and
the overall DM production rate becomes suppressed by $\exp{(-m/T)}$ \cite{Cosme:2023xpa,Bernal:2025fcl}.
Consequently, to satisfy the observed DM density, one needs to assume stronger DM couplings compared to the vanilla freeze-in DM.
This has a striking consequence in the detection aspects, making freeze-in DM discoverable in the next generation experiments while opening up a new viable parameter space for DM genesis, bridging the gap between the parameter spaces for freeze-out and vanilla freeze-in.
In the past few years, several studies have explored this possibility for different models \cite{Silva-Malpartida:2023yks,Koutroulis:2023fgp,Khan:2025keb,Arcadi:2024wwg,Boddy:2024vgt,Arcadi:2024obp,Belanger:2024yoj,Arias:2025tvd,Bernal:2026clv,Cerdeno:2026lgw,Bernal:2026vbg}.

Motivated by these developments, in this work we investigate the low-reheating-
temperature production of vector dark matter in the custodially stabilized
hidden $(SU(2)_{\rm HS})$ model.
We emphasize that neither Boltzmann-suppressed
freeze-in at stronger coupling nor its application to spin-1 dark matter is by
itself new. The possibility of realizing freeze-in with substantially enhanced
couplings when the maximum temperature of the thermal bath lies below the DM
mass was discussed in Ref.~\cite{Cosme:2023xpa}, and was subsequently studied systematically
for Higgs-portal dark matter of spin (0), (1/2), and (1) in ~\cite{Arcadi:2024wwg}.
A fully renormalizable Higgs-portal vector-DM realization based on a
spontaneously broken Abelian $(U(1)_D)$ gauge symmetry at low reheating
temperature was investigated in \cite{Khan:2025keb}. Freeze-in 
production of non-Abelian vector dark matter has been previously considered in a
different $(SU(2)_N)$ (where $N$ stands for neutral) gauge group \cite{Barman:2019lvm}.That construction, however, has a different particle 
content and stabilization mechanism and receives an important freeze-in 
contribution from decays of heavier scalar states; it is therefore distinct 
from the Boltzmann-suppressed low- $T_{\rm RH}$ Higgs-portal production considered here.
The custodially stabilized hidden-$(SU(2))$ vector-DM framework employed here, including its three mass-degenerate gauge bosons and characteristic non-Abelian
interactions, is itself well established \cite{Hambye:2008bq,Baouche:2021wwa,Choi:2017zww}. The aim of the
present work is therefore not to introduce a new freeze-in mechanism
or a new vector-DM model, but rather to determine quantitatively how the
Boltzmann-suppressed low-$(T_{\rm RH})$ nonthermal regime is realized in this
particular three-component custodial $(SU(2)_{\rm HS})$ theory. In contrast to
a single-component Abelian vector realization, spontaneous breaking of
$(SU(2)_{\rm HS})$ leaves a residual custodial $(SO(3))$ symmetry that renders
the three degenerate vector states $(X_i,(i=1,2,3))$ stable without imposing
an additional stabilizing discrete symmetry, and the non-Abelian gauge
structure generates cross-species semi-annihilation reactions
$(X_iX_j\leftrightarrow X_k h)$ and $(X_iX_j\leftrightarrow X_k\eta)$,
with $(i,j,k)$ all distinct. In our analysis, we consistently include
these genuinely non-Abelian processes together with the ordinary Higgs-portal
production channels, the mixed $(h-\eta)$ initial state, and the corresponding
reverse reactions. We explicitly determine the thermalization boundary
beyond which the vector sector can no longer be treated within the freeze-in
regime, and restrict the relic-density analysis to the nonthermal region.
Although we find that semi-annihilation produces only a small numerical
modification of the final abundance over the parameter region of interest,
the threefold custodial degeneracy
implies $(Y_{\rm DM}=Y_{X_1}+Y_{X_2}+Y_{X_3}=3Y_X)$. 
Since the observed relic density constrains the total abundance $ Y_{\rm DM} = 3Y_X$, each of the three degenerate vector DM species needs to account for only 
one third of the required yield. As a result, for otherwise comparable production dynamics, the gauge coupling required to reproduce the observed relic abundance is smaller than in a single-component vector DM scenario (approximately by a factor 
$1/\sqrt{3} $ when $Y_X \propto g^2_\phi $). When this multiplicity effect is combined
with the larger couplings required to compensate the Boltzmann suppression for 
$T_{\rm RH} < m_X $, this leads to a quantitatively distinct direct-detection phenomenology and to regions of the nonthermal parameter space that can already be
tested by LZ and PandaX-4T and further probed by future experiments such
as DARWIN\footnote{For simplicity in this work we restrict to the instantaneous reheating approximation with $Y_\chi(T_{\rm RH})=0$. A detailed analysis with noninstantaneous reheating can be found in ref.\cite{Khan:2025keb,Silva-Malpartida:2023yks}}.

This paper is organized as follows. In Sec.\ref{sec:model} we present the particle content and the basic setup.
Sec.\ref{sec:dmprod} contains the details of obtaining the produced DM density and its dependencies on various model parameters.
We discuss the theoretical and experimental constraints along with our numerical results in Sec.\ref{sec:result}.
Finally, we conclude in Sec.\ref{sec:concl}. Some relevant discussions about the scalar sector have been made in Appendix \ref{sec:apxA}, \ref{sec:apxB}, and \ref{sec:apxC}.

\section{Model}
\label{sec:model}
We consider an extension of the Standard Model (SM) by a non-abelian gauge symmetry SU(2)$_{\rm HS}$, under which the SM particles are singlets. 
The SM particle content is also augmented by a BSM scalar doublet $\phi$, charged under this group. The DM candidates are the three gauge bosons of this group, represented by $X_i ~(i=1,2,3)$. 
The DM can interact with the SM particles through a Higgs-portal interaction. The Lagrangian of this model is written as, \cite{Hambye:2008bq}:
%

\begin{equation}
{\cal L}_{\rm BSM}\supset-\frac{1}{4}V^{a \mu\nu}V^a_{\mu\nu}+(D_{\mu}\phi)^{\dagger}(D^{\mu}\phi)-V(H, \phi)
\label{eq:Lagrangian_model}
\end{equation}
where $ D^{\mu}\phi=\big(\partial^{\mu}-ig_{\phi}\frac{\tau^a}{2} X^{a \mu}\big)\phi$
with $g_{\phi}$ is the $SU(2)_{\rm HS}$ gauge coupling; and $\tau^a$ are the Pauli matrices.
The field tensor of the BSM gauge boson is depicted by,
$V^a_{\mu \nu}= \partial_\mu X^{a}_\nu-\partial_\nu X^{a}_\mu+g_\phi \epsilon^{abc}X^{b}_\mu X^{c}_\nu$, where the $SU(2)$ indices $a,b,c$ run from $1-3$. $H$ represents the SM Higgs doublet charged under the SM gauge group.
The scalar sector of the framework reads as,
\begin{equation}
    V(H,\phi )=\mu_H^2 (H^\dagger H)+\lambda_{H}(H^\dagger H)^2 + \mu_{\phi}^{2}\phi^{\dagger}\phi+\lambda_{\phi}(\phi^{\dagger}\phi)^{2}+\lambda_{\phi H}\phi^{\dagger}\phi H^{\dagger}H
\end{equation}
%
For the spontaneous symmetry breaking in both sectors (SM \& $SU(2)_{\rm HS}$) one requires $\mu_H^2 < 0$ and $\mu_{\phi}^{2}<0$. In the unitary gauge one can write $H$ and $\phi$ as  
\begin{equation}
    H= \begin{pmatrix}
    0\\
    \frac{v+S }{\sqrt{2}}
    \end{pmatrix},~~~~~~~~ 
     \phi= \begin{pmatrix}
    0\\
    \frac{v_\phi+\phi_0 }{\sqrt{2}}
    \end{pmatrix}. 
\end{equation}
where $S$ and $\phi_0$ are physical scalars after the SSB in both the SM and hidden sectors, respectively. $v$ and $v_\phi$ denote the vacuum expectation values along the SM Higgs and hidden-scalar directions, respectively.
Thus, the hidden sector gauge boson $X $ becomes massive by absorbing the Goldstone degrees
of freedom of the $SU(2)_{\rm HS}$ Higgs field $\phi$ and the corresponding Lagrangian is given by,
\begin{eqnarray}
    \nonumber {\cal L}_{\rm int}&\supset&-\frac{1}{4}V^{a \mu\nu}V^a_{\mu\nu}+ \frac{1}{2}(\frac{1}{2}g_\phi v_\phi)^2 X^{a}_\mu X^{ a\mu}+
      \frac{1}{4} g_\phi^2 v_\phi  X^{a}_\mu X^{ a\mu} \phi_0+ \frac{1}{8} g_\phi^2 X_\mu X^\mu \phi_0^2\\
      &&- \frac{1}{2}\lambda_{\phi H}(v_\phi+\phi_0)^2 H^{\dagger}H+\mathcal{L}_{\rm \phi_0},
      \label{eq:lagrangian_full}
      \end{eqnarray}
      where, 
      \begin{eqnarray}
      {\cal L}_{\phi_0} = \frac{1}{2} \left (\partial_\mu \phi_0 \right) \left (\partial^\mu \phi_0 \right ) -\frac{1}{2} \mu_\phi^2 \left (v_\phi + \phi_0 \right)^2 - \frac{1}{4} \lambda_\phi \left (v_\phi + \phi_0 \right)^4.
\end{eqnarray}
The spontaneous breaking of \(SU(2)_{\rm HS}\) has several important consequences. 
First, all three hidden gauge bosons acquire a common mass, $m_{X}=\frac{1}{2}g_\phi v_\phi$ .
Second, the symmetry-breaking pattern leaves an accidental custodial \(SO(3)\) symmetry acting on the triplet of hidden gauge bosons \(X_\mu^{ a}\), with \(a=1,2,3\). 
As a result, the three vector states remain exactly mass-degenerate. 
Moreover, the custodial \(SO(3)\) invariance implies that the hidden gauge bosons enter the scalar interaction terms only through \(SO(3)\)-singlet combinations such as $X_\mu^{ a}X^{ a\mu}$.
Consequently, no interaction vertex involving a single \(X_\mu^{ a}\) is generated after symmetry breaking, and the lightest hidden vector states are stable. 
Thus, the three components \(X_\mu^{ 1,2,3}\) collectively constitute a triply degenerate vector dark matter sector, with identical masses and portal interactions. 
This stability arises dynamically from the residual custodial symmetry and does not require the imposition of an additional ad hoc \(\mathcal {Z} _ 2 \) symmetry, unlike many Abelian vector dark matter constructions.
For example, $U(1)_X$ abelian VDM model would require the BSM gauge boson and a BSM singlet scalar to transform non-trivially under this  $\mathcal Z_2$, while all the SM fields transform trivially under the same $\mathcal Z_2$ ~\cite{Khan:2025keb}. Such a symmetry forbids the kinetic mixing of the $U(1)_X$ vector boson and the $U(1)_Y$ gauge boson, which is crucial to make the DM stable.

After electroweak symmetry breaking (EWSB), the CP-even scalar component of the SM Higgs doublet, $S$, mixes with the CP-even hidden-sector scalar $(\phi_0)$.
The minimization condition of the entire potential i.e. $\partial V(H,\phi)/\partial \phi_0=0=\partial V(H,\phi)/\partial S$ leads to,
\begin{eqnarray}
    \mu_H^2=-\lambda_H v^2-\frac{1}{2}\lambda_{\phi H} v_\phi^2,~~~~ \mu_\phi^2=-\lambda_\phi v_\phi^2-\frac{1}{2}\lambda_{\phi H} v^2
\end{eqnarray}
Thus, the mass matrix of the CP even scalars $(S~~\phi_0)^T$ is given by, 
\begin{eqnarray}
    M^2=\begin{pmatrix}
        2 \lambda_H v^2 &~~ \lambda_{\phi H} ~v v_\phi \\
    \lambda_{\phi H} ~v v_\phi  & 2 \lambda_\phi v_\phi^2 
    \end{pmatrix}
\end{eqnarray}
Upon diagonalization the physical scalar ($h_1,h_2$) masses are given by,
\begin{eqnarray}
m_{h_1,h_2}^{2}=\lambda_H v^{2}+\lambda_{\phi}v_{\phi}^{2}\mp\sqrt{(\lambda_H v^2 -\lambda_\phi v_\phi^2)^2+\lambda_{\phi H}^2~ v^2 v^2_\phi},
\label{eq:mass}
\end{eqnarray}
$h_1,h_2$ signify the physical $h,\eta$ depending on the mass hierarchy.
In our convention, $h$ represents the SM Higgs boson with mass $m_h = 125 $ GeV. In equation \ref{eq:mass}, $-(+)$ sign corresponds to $m_h (m_\eta)$ for the case $m_h< m_\eta$. Conversely, for $m_h> m_\eta$, corresponding masses are obtained by interchanging the $\pm $ signs in the same equation.
$h,\eta$ is related to the unphysical basis by,
\begin{eqnarray}
    \begin{pmatrix}
        h\\
        \eta
    \end{pmatrix}=
    \begin{pmatrix}
        \cos\beta & -\sin\beta\\
        \sin\beta & \cos\beta
    \end{pmatrix}
    \begin{pmatrix}
        S\\
        \phi_0
    \end{pmatrix},~~~{\rm with,}~~\tan2\beta=\frac{\lambda_{\phi H} v v_\phi}{\lambda_\phi v_\phi^2-\lambda_H v^2}
\end{eqnarray}

\begin{figure}[!tbh]
    \centering
    \includegraphics[scale=0.4]{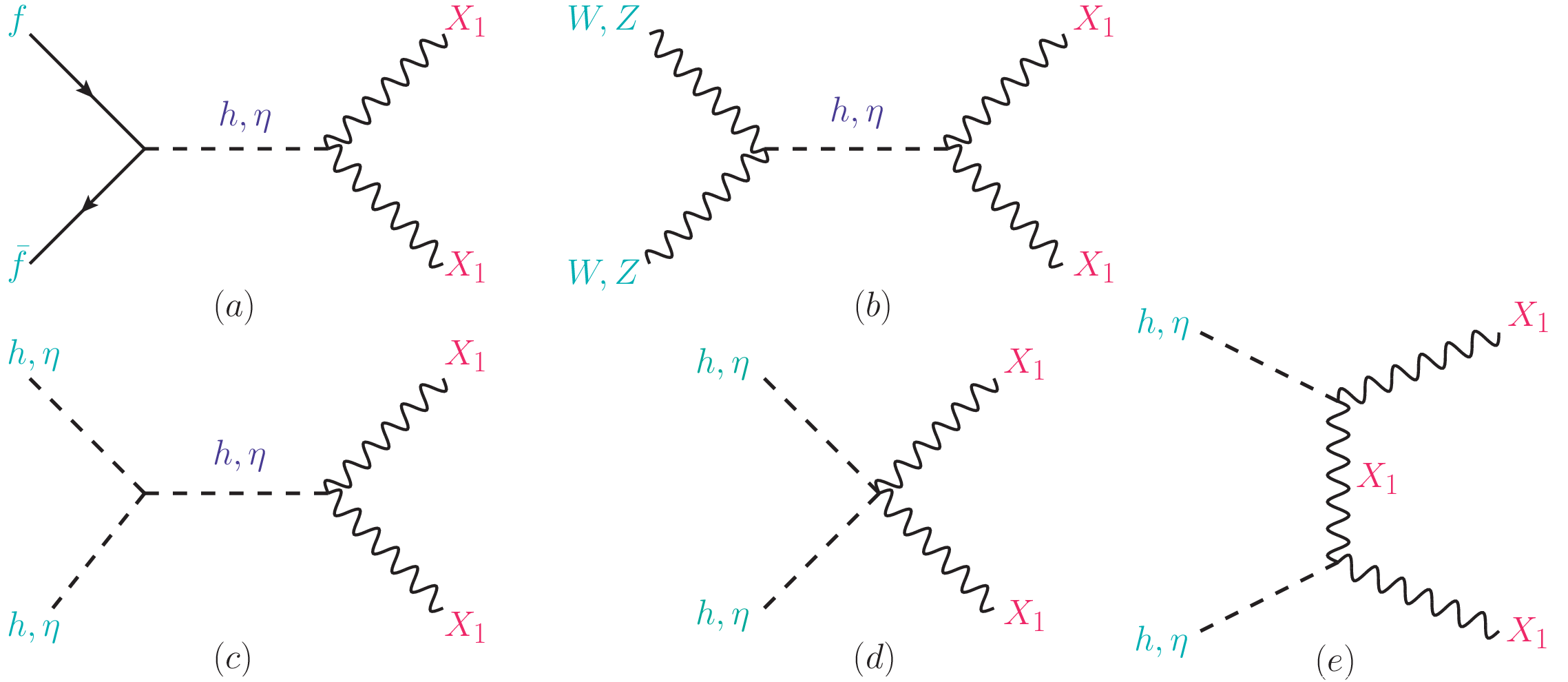}
    \caption{Feynman diagrams relevant for DM production. The channels are (a) fermion annihilation via scalar mediator, (b) gauge boson annihilation via scalar mediator, (c) scalar annihilation via scalar mediator, (d) scalar annihilation w/o mediator, (e) scalar annihilation via VDM mediator. For the last one, there also exists a $u$-channel diagram, which is not shown explicitly.}
    \label{fig:fd}
\end{figure}

Thus, the main interaction channel of $X$ is through the Higgs portal ($h,\eta$) with SM fermions and the SM gauge bosons governed by $\sin2\beta \times (g_\phi v_\phi)$, which is important from both production and detection aspects.
Also there can be quartic vertex with $h,\eta$ governed by  $\sin\beta (\cos\beta) \times g_\phi $ and scalar portal interactions of $hh (\eta \eta)\to XX$ governed by same $\sin\beta$ and $(g_\phi v_\phi)$.
The relevant Feynman diagrams related for  DM production are shown in Fig.\ref{fig:fd}.
The model parameters are the following: 
\begin{eqnarray}
    m_{X},~ g_\phi,~ \sin\beta,~ m_\eta
\end{eqnarray}
Since $X$ is the DM candidate, we use $M_{\rm DM}$ and $m_X$ interchangeably. The other dependent variables in terms of these independent parameters can be found in Appendix \ref{sec:apxA}.


\section{DM production}
\label{sec:dmprod}
Following the detailed discussion of the BSM model in the previous section, we are now set to analyze the freeze in DM production in the early universe.
As mentioned in the Introduction, the reheating temperature $T_{\rm RH}$ is one of the key inputs that governs the freeze-in abundance, which will be shown in this section.
The DM $X$ gets produced from the BSM interactions of SM particles as well as the BSM scalar $\eta$. The evolution of $X$ density can be obtained by solving the Boltzmann equation as a function of temperature $T$,
\begin{eqnarray}
    \frac{dY_{X}}{dT}&=& -\frac{S_{\rm ent}}{HT}\sum_{i={ f,h,\eta,V}} \left[(Y_i^{eq})^2 \langle \sigma v\rangle_{ii\to XX} -(Y_{X})^2 \langle \sigma v\rangle_{XX\to ii}\right]\nonumber\\
    &&
    -\frac{S_{\rm ent}}{HT}
    \left[Y_h^{eq}Y_\eta^{eq} \langle \sigma v\rangle_{h\eta\to XX} -(Y_{X})^2 \langle \sigma v\rangle_{XX\to h\eta}\right]
    \nonumber\\
    &&
    -\frac{S_{\rm ent}}{HT}
    \sum_{i=h,\eta}\langle \sigma v\rangle_{XX\to X i} \left[Y_{X}Y_X^{eq}-(Y_X)^2 \right]
    \nonumber\\
    &&-\frac{1}{HT}\sum_{j={\rm h,\eta}}\left[ (Y_j^{eq} \langle \Gamma\rangle_{j\to XX}-Y_{X} \langle \Gamma\rangle_{XX\to j})\right],
    \label{eq:boltz}
\end{eqnarray}
where, $Y_X$ and  $Y_i^{eq}$ signify the comoving number density of DM $X$ and comoving equilibrium number density of the species $i~(\equiv f,h,V,\eta)$. Moreover, $S_{\rm ent}$ stands for the comoving entropy density, whereas $H=\sqrt{8\pi\rho_{\rm rad}/3}/M_{\rm pl}$ is the Hubble expansion rate with $M_{\rm pl}$ and $\rho_{\rm rad}$ being the Planck mass and the total energy density, respectively.
The initial condition is imposed by setting $Y_{X}=0$ at $T=T_{\rm RH}$.
Note that the total DM abundance $Y_{\rm DM}$ can be obtained by adding the abundances of different DM species i.e. $Y_{\rm DM} =3 Y_{X}$. Here, the multiplication factor 3 implies the three components of $X_i~(i=1,2,3)$.

The right-hand side (RHS) of the aforementioned equation encodes the collision terms with all the relevant interactions as shown in Fig.\ref{fig:fd}.
The term in the first line of the RHS corresponds to the annihilation of SM particles (fermions, gauge bosons, and Higgs) as well as the BSM scalar $\eta$.
The second term in the same first line takes care of the possible back-reaction of $X$ and the subsequent dilution of $X$.
Still, for consistency, we always include the back-reaction processes in our numerical implementation.
However, the second contribution is practically negligible given the very small initial abundance of $X$.
On the other hand, the term in the second line in the RHS contains the contribution from the decay of the scalar particle $j(\equiv h,\eta)$ when $2 m_X\leq m_j$.
The thermal averages of the respective annihilation cross section ($\sigma$) and decay width ($\Gamma$) are denoted by $\langle\sigma v\rangle$ and $\langle\Gamma\rangle$, respectively.
The production cross sections for pair annihilation of SM fermions and gauge bosons are given as,
\begin{eqnarray}
    \sigma_{f\bar f \rightarrow XX} &=& \frac{1}{N_c}\frac{ m_f^2\sin^2 2\beta}{256 \pi sv^2 v_{\phi}^2}\bigg|\frac{1}{s-m_h^2+i m_h\Gamma_h}-\frac{1}{s-m_\eta^2+i m_\eta\Gamma_\eta}\bigg|^2(s^2-4 m_X^2 s+12 m_X^4)\nonumber\\
    &&\sqrt{(s-4 m_f^2)(s-4 m_X^2)}\\
    \sigma_{VV\rightarrow XX}& =&\frac{\sin^2 2\beta}{1152\pi s v^2 v_\phi^2}\bigg|\frac{1}{s-m_h^2+i m_h\Gamma_h}-\frac{1}{s-m_\eta^2+i m_\eta\Gamma_\eta}\bigg|^2\left( s^2-4 m_X^2 s+12 m_X^4\right) \nonumber\\
    &&\times\left( s^2-4 m_V^2 s+12 m_V^4\right)\sqrt{\frac{(s-4 m_X^2)}{(s-4 m_V^2)}}.
\end{eqnarray}
$m_f~(m_V)$ signifies the mass of SM fermion (gauge boson i.e. $Z,W^\pm$) and $N_c$ stands for the color factor.
Note that $N_c=3$ for quarks and $N_c=1$ for leptons.
Apart from these, there also exist scattering processes like $hh~(\eta \eta) \to XX$ and $h\eta\to XX$ which are summarized in Appendix \ref{sec:apxB}.
The third line in eq.\eqref{eq:boltz} signifies the contribution from semi-annihilation process $X_i X_j\to X_k h(X_k\eta)$~($i,j,k$ different) \cite{Choi:2017zww,Baouche:2021wwa} and the corresponding cross-section is summarized in Appendix \ref{sec:apxD}.
Finally, the decay widths are given by,
\begin{eqnarray}
   \Gamma\,(h\rightarrow X_{i}X_{i})&=&\frac{g_{\phi}^{2}\,m_{h}^{3}\sin^2{\beta}}{128 \pi m_{X}^{2}}\sqrt{1-4\frac{m_{X}^{2}}{m_{h}^{2}}}\left\{ 1-4\frac{m_{X}^{2}}{m_{h}^{2}}+12\frac{m_{X}^{4}}{m_{h}^{4}}\right\}\\
   \Gamma\,(\eta\rightarrow X_{i}X_{i})&=&\frac{g_\phi^2 m_\eta^3 \cos^2{\beta}}{128\pi m_{X}^2}\sqrt{1-4\frac{m_{X}^{2}}{m_{\eta}^{2}}}\left\{ 1-4\frac{m_{X}^{2}}{m_{\eta}^{2}}+12\frac{m_{X}^{4}}{m_{\eta}^{4}}\right\}
\end{eqnarray}

\begin{figure}[!tbh]
    \centering
    \subfigure[\label{a1}]{
    \includegraphics[scale=0.4]{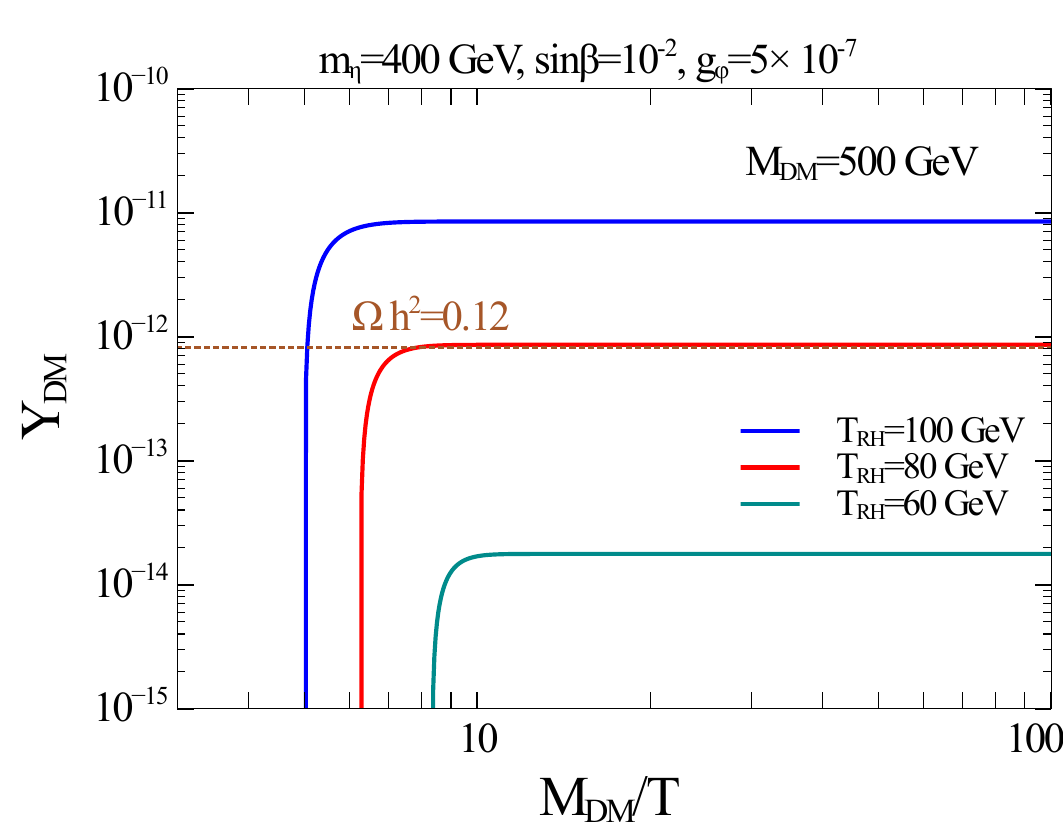}}
    \subfigure[\label{a2}]{
    \includegraphics[scale=0.4]{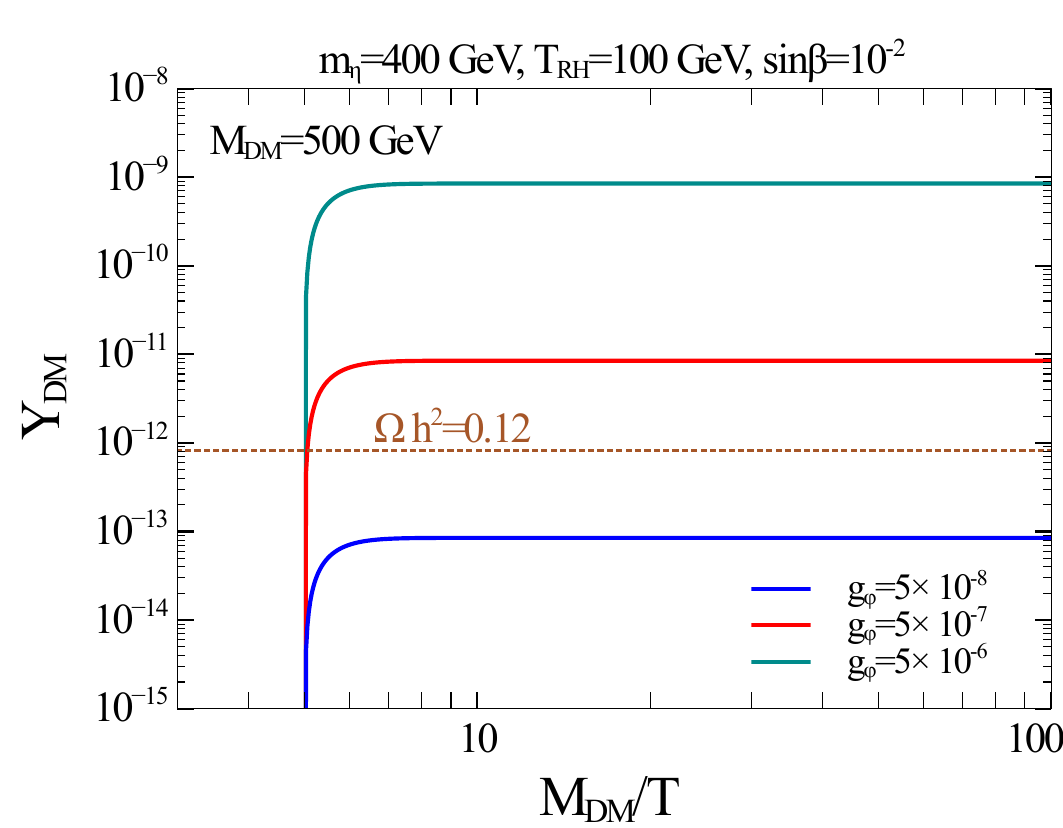}}
    \caption{Evolution of co-moving number densities of DM $Y_{\rm DM}$ with respect to $M_{\rm DM}/T$ for a benchmark $m_\eta=400$ GeV, and $\sin\beta=10^{-2}$. {\bf Left:} Variation of $Y_{\rm DM}$ with different values of $T_{\rm RH}$ for a fixed coupling
    $g_\phi=5\times 10^{-7}$. The cyan, red and blue lines correspond to $T_{\rm RH}=60~{\rm GeV},~80~{\rm GeV}~{\rm and}~100~{\rm GeV}$, respectively.
    {\bf Right:} Variation of $Y_{\rm DM}$ with different values of $g_\phi$ for a fixed 
    $T_{\rm RH}=100$ GeV. The cyan, red and blue lines correspond to $g_\phi= 5\times 10^{-8},~5\times 10^{-7}~{\rm and}~5\times 10^{-6}$, respectively.}
    \label{fig:abund}
\end{figure}

We numerically solve eq.\eqref{eq:boltz} and show the evolution of comoving abundance of DM with respect to $M_{DM}/T$ in Fig.\ref{fig:abund}.
For this plot, we consider a benchmark value of model parameters $m_\eta=400$ GeV, $\sin\beta=10^{-2}$.
In Fig.\ref{a1} we display the 
comoving abundances for different values of $T_{\rm RH}=60~{\rm GeV},~80~{\rm GeV}~{\rm and}~100~{\rm GeV}$ for a fixed value of $g_\phi=5\times 10^{-7}$ depicted by cyan, red, and blue solid lines, respectively.
As expected, starting from a negligible abundance, $X$ promptly gains its freeze-in density around $T_{\rm RH}$.
Note that, with a decrease in $T_{\rm RH}$, the abundance also decreases.
This can be understood from the temperature dependence of the interaction rates encoded in $\langle \sigma v\rangle$ and $\langle \Gamma \rangle$.
For example $\langle \sigma v\rangle$ reads as 
\begin{eqnarray}
   \langle \sigma v\rangle_{ij\to XX} = \frac{g_i g_j T}{2(2\pi)^4~n_{i}^{eq}~n_{j}^{eq}} \int_{{\rm Max}[4 m_{X}^2,( m_i+m_j)^2]}^{\infty} ds~\sigma_{ij\to XX} \dfrac{\lambda(s,m_i^2,m_j^2)}{\sqrt{s}}K_1\left(\frac{\sqrt{s}}{T} \right),
   \label{eq:thermal_sigma}
\end{eqnarray}
where, $\lambda(s,m_i^2,m_j^2)= [s-(m_i+m_j)^2][s-(m_i-m_j)^2]$ and
$i,j$ signify the particles participating in the DM production. $g_i$, $m_i$ and $n_i$ represent the degree of freedom, mass and density of the particle $i$.

Note that, in the limit 
$T\ll m_{X}$, for processes like $ii\to XX$ the integration in the above equation is maximized when $s=4 m_{X}^2$ for $m_X>m_i$. In this limit, the Bessel function also behaves as $
    K_1\left({\sqrt{s}}/{T}\right) \propto \sqrt{{ T}/{s^{1/2}}} e^{-\sqrt{s}/T}$.
Thus, with an increase in the ratio $m_X/T_{\rm RH}$, the interaction rate and consequently $Y_{\rm DM}$ become exponentially suppressed as observed in Fig.\ref{a1}.
This maps the physical scenario where the interaction rate samples only the high-energy tail of the thermal distribution of bath particles as $T_{\rm RH}$ decreases, leading to a decrease in the yield.
In Fig.\ref{a2} we showcase the variation 
$Y_{\rm DM}$ with different values of $g_\phi$ for a fixed $T_{\rm RH}=100$ GeV,
the cyan, red, and blue lines correspond to $g_\phi= 5\times 10^{-8},~5\times 10^{-7}~{\rm and}~5\times 10^{-6}$.
Needless to say, with an increase in the coupling $g_\phi$, the relevant collision terms increase as $\sim g_\phi^2$, leading to an enhanced density of DM.
In other words, it can be stated that to satisfy the same DM abundance, the associated DM couplings should be higher for a lower value of $T_{\rm RH}$.
This crucial feature serves as the primary focus of this work.
It is also worth highlighting that even with the same value of associated DM cross-section, the obtained DM abundance $Y_{\rm DM}$ in this model is higher (by a factor $ 3$) than other Higgs portal $U(1)_X$ VDM models \cite{Khan:2025keb}. This feature helps to explore a new parameter space with new phenomenological aspects, as will be shown later on.

\begin{figure}[!tbh]
    \centering
    \subfigure[\label{b1}]{
    \includegraphics[scale=0.4]{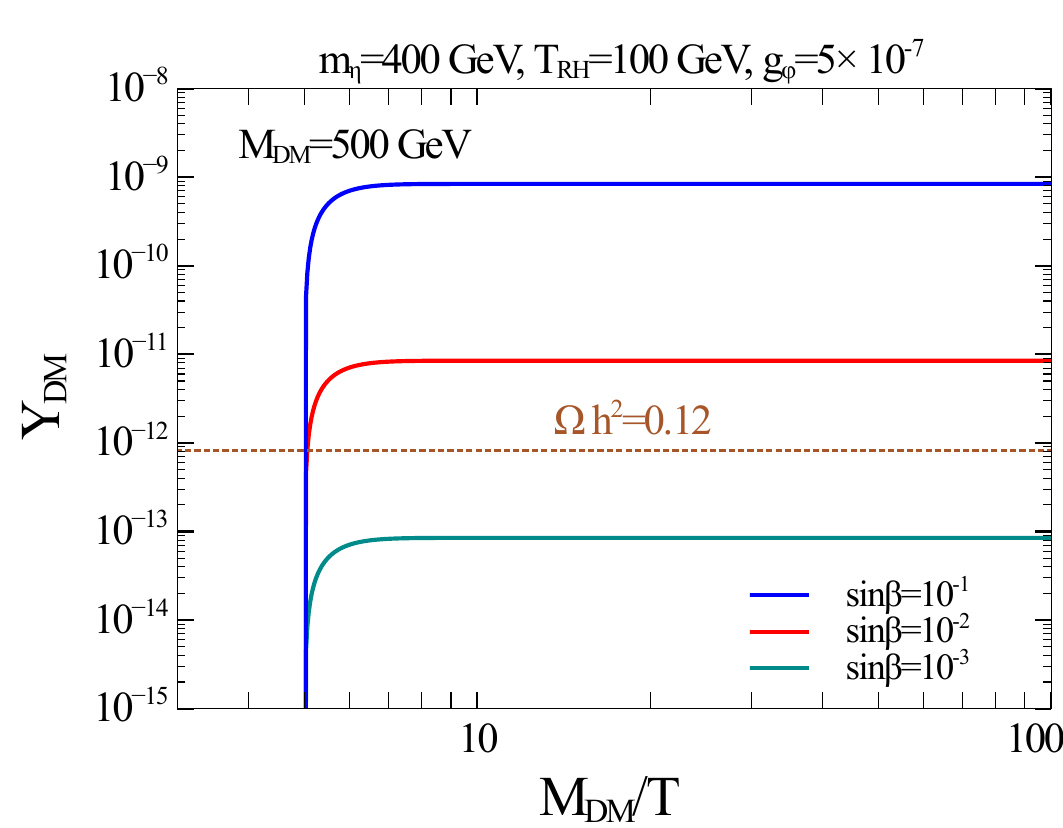}}
    \subfigure[\label{b2}]{
    \includegraphics[scale=0.4]{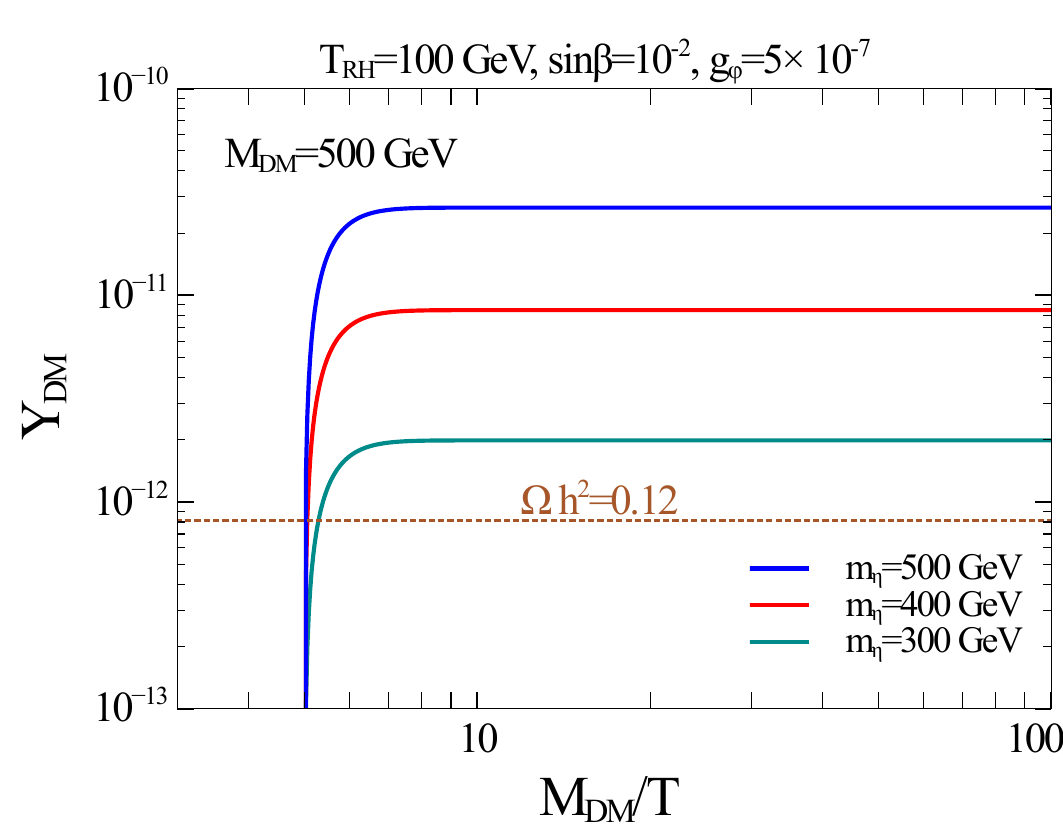}}
    \caption{Evolution of co-moving number densities of DM $Y_{\rm DM}$ with respect to $M_{\rm DM}/T$ for a benchmark  $T_{\rm RH}=100$ GeV and $g_\phi= 5\times 10^{-7}$.
    {\bf Left:} Variation of $Y_{\rm DM}$ with different values of $\sin\beta=10^{-1},~10^{-2},~10^{-3}$
    depicted by cyan, red, and blue lines, respectively, for a fixed $m_\eta=400$ GeV.
    {\bf Right:} Variation of $Y_{\rm DM}$ with different values of $m_\eta$ for a fixed $\sin\beta=10^{-2}$. The cyan, red and blue lines correspond to $m_\eta=300~{\rm GeV},~400~{\rm GeV}~{\rm and}~500~{\rm GeV}$, respectively.}
    \label{fig:abund2}
\end{figure}

We analyze the dependence of DM density on other model parameters in Fig.\ref{fig:abund2}.
For this plot, we now take the fixed values of  $T_{\rm RH}=100$ GeV and $g_\phi= 5\times 10^{-7}$.
In Fig.\ref{b1} we display the 
comoving abundances for different values of $\sin\beta=10^{-1},~10^{-2},~10^{-3}$ for a fixed value of $m_\eta=400$ GeV depicted by cyan, red, and blue solid lines, respectively.
One can observe that the DM abundance increases with $\sin\beta$ as the cross-section has explicit dependence on it, which essentially enhances the collision terms that generate the DM density. 
On the other hand, in Fig.\ref{b2} we showcase the 
comoving abundances for different values of $m_\eta=300~{\rm GeV},~400~{\rm GeV}~{\rm and}~500~{\rm GeV}$  for a fixed value of $\sin\beta=10^{-2}$ depicted by cyan, red, and blue solid lines, respectively.
Here, one can surmise that higher values of $m_\eta$ correspond to a higher abundance of DM.
This trend should
be understood as a result of the combined \(m_\eta\)-dependence
of the scalar cross-section/decay, mediator propagators, interference effects,
and thermal phase-space integrals. Indeed the scalar-annihilation channels,
such as \(\eta\eta\to X_iX_i\) and \(h\eta\to X_iX_i\),
increasing \(m_\eta\) can increase the typical 
\(\sqrt{s}\) and hence leads to a Boltzmann suppression 
through \(K_1(\sqrt{s}/T)\). 
However, the growth of $Y_{\rm DM}$ with $m_\eta$    
is not a purely kinematic phase-space effect, but rather a 
benchmark-dependent result due to the interplay among  scalar amplitudes 
and thermal suppression factors.

After having a detailed discussion on the dependencies of DM density on various model parameters, we will discuss the phenomenological aspects of such a setup through comprehensive numerical analysis.

\section{Results and constraints}
\label{sec:result}
In the previous section, we argued that a low $T_{\rm RH}(<m_{\rm DM})$  requires a higher value of DM coupling to satisfy the observed relic density.
In this section, we perform exhaustive numerical scans over various model parameters
to understand phenomenological implications.
However, the BSM extension of a singlet scalar $\phi$ and three $X_i$ particles can have important implications on various existing
experimental data.
Hence, to ensure phenomenological viability, it is essential to critically examine the aforementioned BSM scenario in light of existing experimental data.
In addition to that, theoretical consistency imposes additional constraints, requiring the model parameters to satisfy a set of well-defined conditions.
Before delving into the numerical analysis, we briefly discuss the existing constraints on the model parameters.

\paragraph{Theoretical constraints:}
The model parameters are constrained by perturbative unitarity, perturbativity, and vacuum stability.
The unitarity of scalar scattering amplitudes imposes the following conditions:
\begin{align}
    \mid \lambda_{\phi H} \mid  &\leq 8\pi\,, \qquad
    \mid \lambda_H \mid \leq 4\pi\,, \qquad
    \mid \lambda_\phi \mid \leq 4\pi\,, \\
    \mid 3(\lambda_H &+ \lambda_\phi) \pm \sqrt{9(\lambda_{H} - \lambda_\phi)^2 + 4\lambda_{\phi H}^2} \mid \leq 8\pi\,.
\end{align}
Furthermore, perturbativity and the requirement of the scalar potential to be bounded from below further imply,
\begin{equation}
    \lambda_H > 0\,, \qquad \lambda_\phi > 0\, , \qquad \lambda_{\phi H}> -2\sqrt{\lambda_H\lambda_\phi}\,.
\end{equation}
The quartic couplings as a function of the scalar masses can be found in Appendix \ref{sec:apxA}.
The perturbative limit on the gauge coupling imposes $g_\phi< \sqrt{4\pi}$.

\paragraph{Collider bounds:}
In the present setup, the SM Higgs boson couples to both the DM $X$ and the other physical neutral scalar $\eta$ ( eq.~\eqref{eq:lagrangian_full}). 
Thus, in the kinematically allowed mass hierarchy, the SM Higgs can decay to these BSM channels, contributing to the invisible decay.
Hence, one very important bound can come from the measurements of the
invisible decay width of the SM Higgs.
For $m_h>2 m_{X}$, the Higgs can decay invisibly into DM particles, with total width
\begin{equation}
    \Gamma_{X} = \sum_{i=1}^{3} \Gamma(h \to X_i X_i)
    = 3\,\Theta(m_h - 2m_X)\,
    \frac{g_\phi^2 m_h^3 \sin^2\beta}{128\pi m_X^2}
    \sqrt{1 - \frac{4m_X^2}{m_h^2}}
    \left( 1 - \frac{4m_X^2}{m_h^2} + 12\frac{m_X^4}{m_h^4} \right).
\end{equation}
The factor of three accounts for the three mass-degenerate $SU(2)_{\rm HS}$ gauge 
bosons. If $m_h > 2 m_\eta $, the Higgs can decay into $\eta$ pairs,
\begin{equation}
    \Gamma(h \to \eta\eta)
    = \Theta(m_h - 2m_\eta)\,
    \frac{\rho_1^2}{32\pi m_h}
    \sqrt{1 - \frac{4m_\eta^2}{m_h^2}}\,.
\end{equation}
The total Higgs decay width is then
\begin{equation}
    \Gamma^{\rm total}_h = \cos^2\beta\, \Gamma_h^{\rm SM} + \Gamma_{XX} + \Gamma_{h\to \eta \eta}
\end{equation}
where $\Gamma_h^{\rm SM} = 4.2~\text{MeV}$ is the total decay width of the Higgs boson into SM particles~\cite{ParticleDataGroup:2024cfk}. 
If $m_\eta> 2 m_X$, then $\eta$ will dominantly decay invisibly, otherwise
$\eta$ will decay back to SM particles through mixing.
Thus, the total invisible decay of the Higgs is given by,
\begin{eqnarray}
    \Gamma_{\rm inv}= \Gamma_X + \Gamma(h\to \eta \eta)\times {\rm BR}( \eta \to XX)^2
\end{eqnarray}
and the corresponding ${\rm BR}_{\rm inv}$ is defined as:
\begin{equation}
    \mathcal{B}_{\rm inv}
    = \frac{ \Gamma_{\rm inv}}{\Gamma^{\rm total}_h}
\end{equation}
For $m_\eta > 2 m_X$, the 
The branching fraction into invisible is constrained to be \cite{CMS:2026nce} 
\begin{eqnarray}
  \mathcal{B}_{\rm inv} \leq 0.13. 
\end{eqnarray}
The CMS Run-2 Higgs combination~\cite{CMS:2026nce} also distinguishes the
branching fraction of the Higgs boson into invisible final states,
$\mathcal{B}_{\rm inv}$, from that into undetected non-standard final
states, $\mathcal{B}_{\rm undet}$. For our model, 
\begin{eqnarray}
\mathcal{B}^{\rm model}_{\rm undet} = \frac{\Gamma(h \to \eta \eta)}{\Gamma^{\rm total}_{h}}; ({\rm for}~m_\eta < 2 m_X)
\end{eqnarray}
In the effective-coupling fit where
both $\mathcal{B}_{\rm inv}$ and $\mathcal{B}_{\rm undet}$ are allowed
to vary, with the additional requirement on the SM $h-WW$ and $h-ZZ$ couplings
modifiers $|\kappa_W|,|\kappa_Z|\leq 1$, CMS obtains the observed 95\% C.L.
limits
\begin{equation}
    \mathcal{B}_{\rm inv}<0.13,
    \qquad
    \mathcal{B}^{\rm model}_{\rm undet}<0.21 .
\end{equation}
The condition $\mid \kappa_W\mid, \mid \kappa_Z \mid \leq1$ is automatically satisfied here, because $\kappa_W=\kappa_Z=\cos\beta\leq1$.
We therefore adopt these limits together with the total-width requirement 
\begin{equation}
    1.0~\text{MeV} < \Gamma_h < 6.0~\text{MeV}\,.
\end{equation}
These conditions impose constraints on $g_\phi$ and $\sin\beta$ for a benchmark $m_{X}$ and $m_\eta$.
Apart from the invisible decay, Higgs signal strength measurements impose a weaker constraint on the mixing angle, which restricts $\sin{\beta}<0.23$ \cite{ATLAS:2016neq}.
Among these, the invisible decay is found to set the strongest constraint.

%
%

\paragraph{Direct detection:}
A direct detection experiment can potentially provide the direct signal of an incoming DM.
While the Earth passes through the galactic DM halo, DM particles are expected to scatter target nucleons in Xenon-based detectors, providing a distinct recoil signal.
Null observation of any such recoil signal stringently constrains the DM nucleon elastic scattering cross-section.
The DM candidate $X$ interacts with nucleons via scalar mediation through the Higgs boson $h$ and the scalar $\eta$. The resulting spin-independent (SI) scattering cross section is given by
\begin{equation}
    \sigma_{\rm SI}
    = \frac{1}{64\pi}\, f^2 m_N^2
    \frac{v_\phi^2}{v^2} g_\phi^4 \sin^2(2\beta)\,
    \frac{(m_\eta^2 - m_h^2)^2}{m_\eta^4 m_h^4}\,
    \frac{\mu_r^2}{m_X^2}\,,
\end{equation}
where $\mu_r = \frac{m_N m_X}{m_N + m_X}$ denotes the reduced mass of the DM--nucleon system, and $f$ parametrizes the effective Higgs--nucleon coupling given as,
\begin{eqnarray}
   f\equiv \frac{\langle N | \sum_q \bar{q}q| N \rangle}{m_N}
\end{eqnarray}

\begin{figure}[!tbh]
 \centering
 \subfigure[]{\includegraphics[scale=0.4]{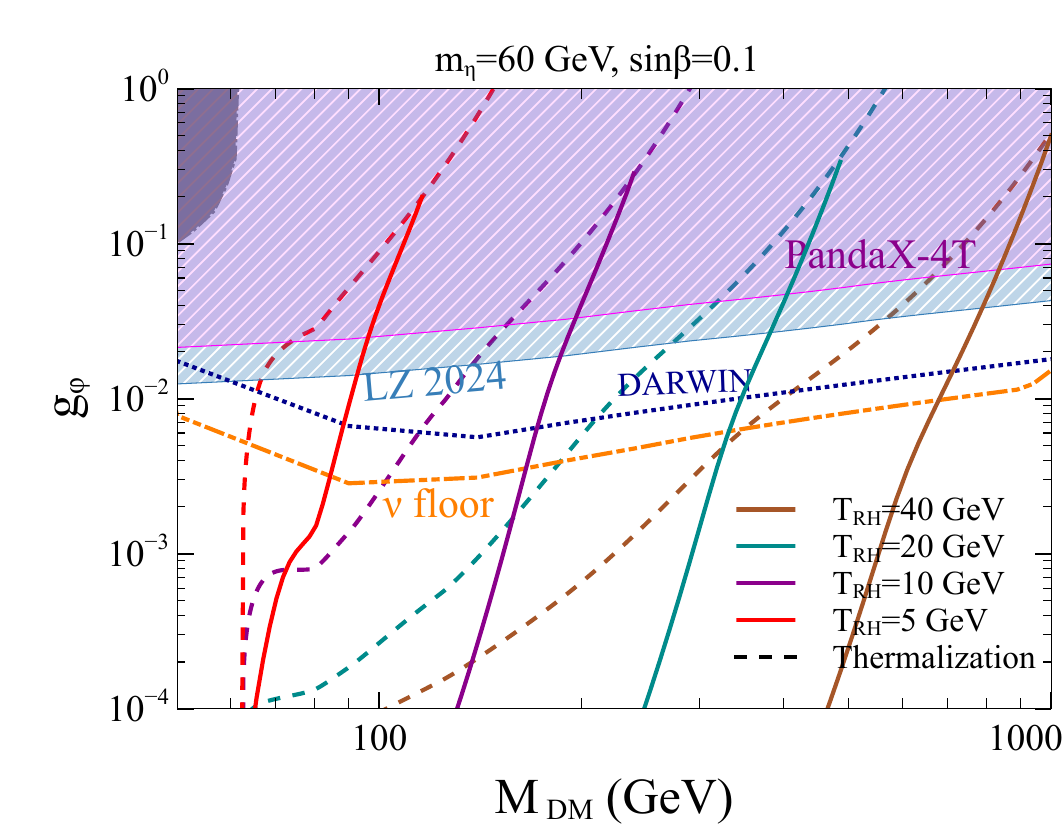}}
\subfigure[]{\includegraphics[scale=0.4]{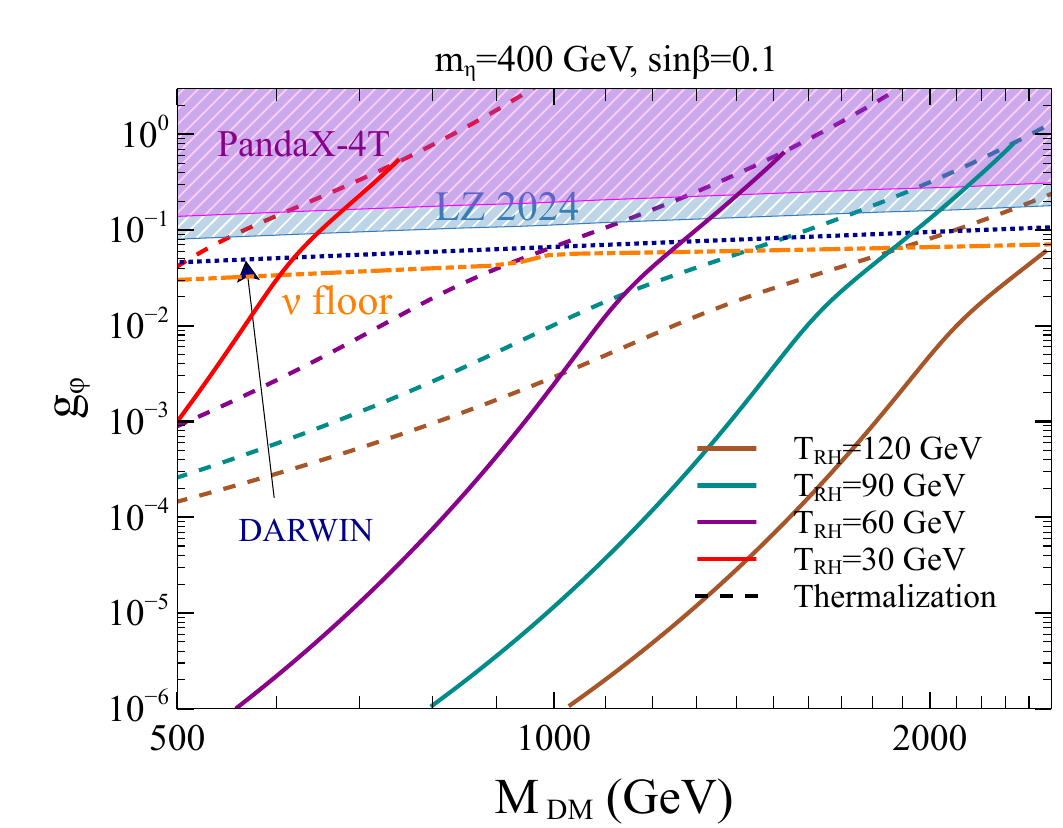}}
    \caption{Contours satisfying the observed relic density are shown in the $g_\phi$--$M_{\rm DM}$ plane. Different contours correspond to varying reheating temperatures $T_{\rm RH}$. The direct detection limits from the LZ 2024, PandaX-4T experiments are overlaid as exclusion bounds. The panel $(a)$ and $(b)$ correspond to $m_\eta = 60~\text{GeV}$ and $400~\text{GeV}$ respectively. In both cases, the scalar mixing angle is fixed to $\sin\beta = 0.1$.
    The dashed colored lines signify the thermalization boundaries for different values of $T_{\rm RH}$. The gray shaded region in panel (a) signifies the excluded parameter space from the Higgs invisible decay constraint.}
    \label{fig:direct_detection}
\end{figure}

We take the value $f=0.3$.
For this work, we consider the existing constraints from experiments like LZ \cite{LZ:2024zvo}, PandaX-4T \cite{PandaX-4T:2021bab}
(Fig.~\ref{fig:direct_detection}).

\begin{figure}[!tbh]
 \centering
 \subfigure[]{\includegraphics[scale=0.4]{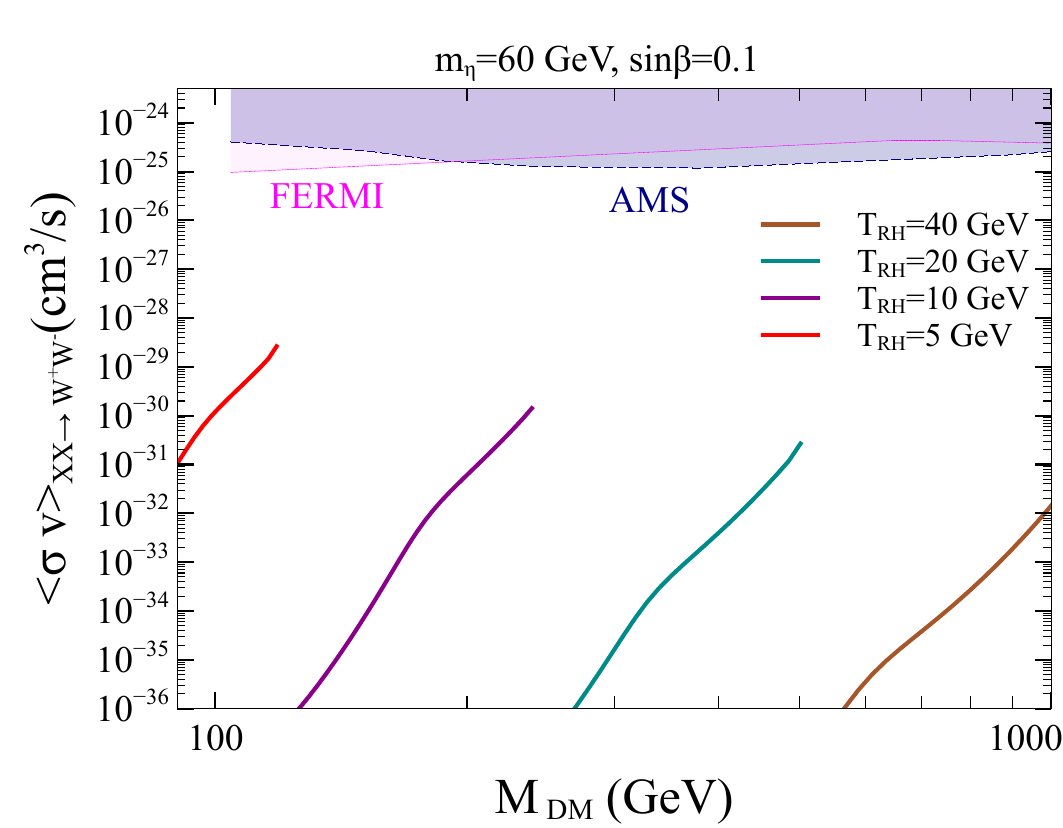}}
\subfigure[]{\includegraphics[scale=0.4]{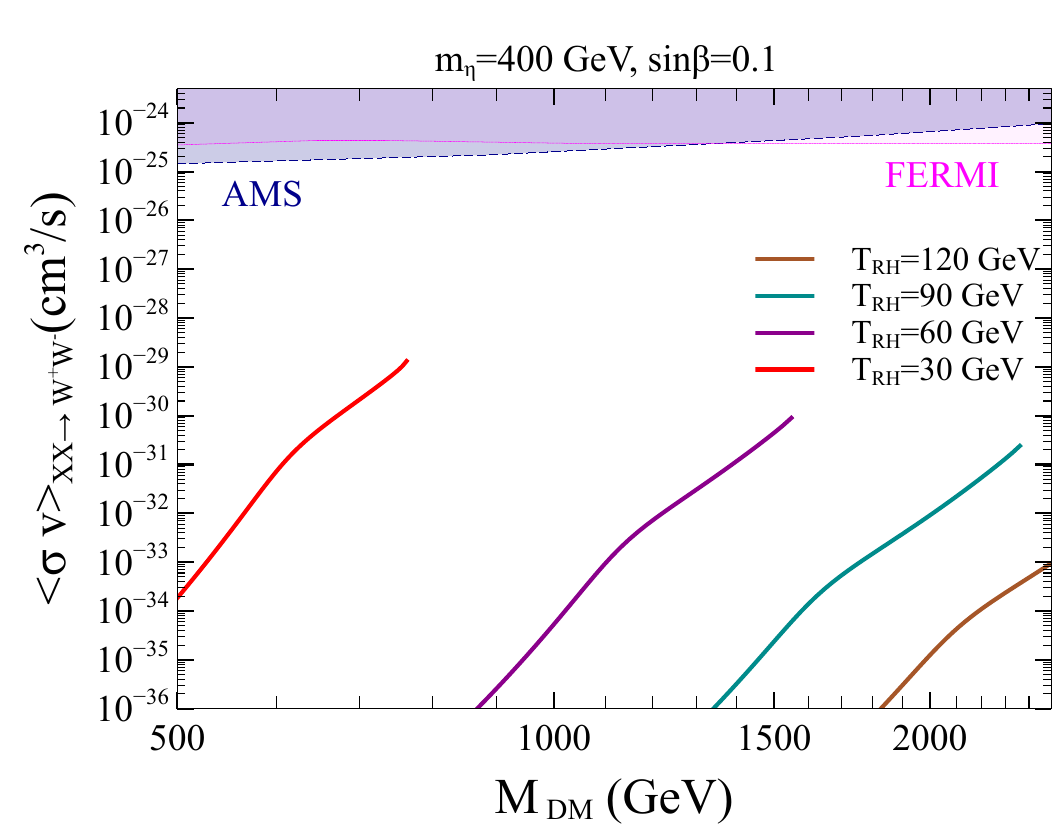}}
    \caption{Contours satisfying the observed relic density are shown in the thermal average annihilation cross-section ($\langle \sigma v\rangle$) of DM to $W^+W^-$ vs. $M_{\rm DM}$ plane, with the different lines corresponding to the different $T_{\rm RH}$, similar to figure~\ref{fig:direct_detection}. Indirect detection bounds from the \textit{Fermi}-LAT and AMS experiments are shown. The left and right panels correspond to the same parameter choices as in figure~\ref{fig:direct_detection}.}
    \label{fig:indirect_detection}
\end{figure}

\paragraph{Indirect detection:}
DM can pair annihilate into SM particles via $h$ and $\eta$ portal interactions. 
The null observation of any excess signal in different space-based or ground-based telescopes strongly constrains such annihilation channels.
For our analysis, the relevant indirect-detection constraint arises from DM annihilation into gauge bosons and quarks.
We discuss such bounds on $X_iX_i\rightarrow W^+W^-$ channel from the FERMI-LAT \cite{Fermi-LAT:2015att} and AMS \cite{Calore:2022stf} experiments, which are found to be the strongest (Fig. \ref{fig:indirect_detection}). 

\paragraph{Results:}We present contour plots satisfying the observed DM relic abundance in the $M_{\rm DM}$--$g_\phi$ plane in figure~\ref{fig:direct_detection}. 
The DM relic abundance is estimated from $Y_{\rm DM}$,
\begin{eqnarray}
    \Omega h^2=2.755\times10^8~ Y_{\rm DM}~ (m_{\rm DM}/{\rm GeV}).
\end{eqnarray}
We vary the DM mass and coupling for two benchmark mediator masses, $m_\eta = 60~\text{GeV}$ ({\bf left}) and $400~\text{GeV}$ (({\bf right})), and fixed mixing angle $\sin\beta = 0.1$.
Note that the chosen benchmark is consistent with theoretical constraints and collider bounds \cite{Baouche:2021wwa}.
The different colored solid lines correspond to different values of $T_{\rm RH}$ that satisfy the observed relic density $\Omega h^2=0.12$ \cite{Planck:2018vyg}.
In the left plot, the red, dark magenta, cyan, and brown solid lines signify $T_{\rm RH}=5$ GeV, 10 GeV, 20 GeV, and 40 GeV, respectively.
On the other hand, in the right plot, the red, dark magenta, cyan, and brown solid lines signify $T_{\rm RH}=30$ GeV, 60 GeV, 90 GeV,  and 120 GeV, respectively.

In the plots, we also highlight the thermalization boundaries by colored dashed lines corresponding to the different values of $T_{\rm RH}$.
The thermalization boundary is obtained by imposing 
\begin{eqnarray}
    \Gamma_X(T_{\rm RH}) \leq H(T_{\rm RH}),
\end{eqnarray}
where, $\Gamma_X(T)$ is the interaction rate given by,
\begin{eqnarray}
    \Gamma_X &=& \left(n_X^{\rm eq}\right)^{-1}\Bigg[\sum_{i={ f,h,\eta,V}} n_i^{eq} n_i^{eq}\langle \sigma v\rangle_{ii\to XX}
    +
    n_\eta^{eq} n_h^{eq} \langle \sigma v\rangle_{h\eta\to XX} \Bigg]
    \nonumber\\
    &&
    +
    \sum_{i=h,\eta}  n_i^{\rm eq} \langle \sigma v\rangle_{X i\to X X}
    +\left(n_X^{\rm eq}\right)^{-1}\sum_{j={\rm h,\eta}}n_{j}^{\rm eq}\langle\Gamma\rangle_{j\to XX}.
    \label{eq:rate}
\end{eqnarray}
Note that we work with the approximation of instantaneous reheating, treating $T_{\rm RH}$ as the maximum temperature.
In the parameter space above those dashed lines, DM eventually gets thermalized for the given model parameters and $T_{\rm RH}$.
We restrict our analysis of freeze-in DM below those boundaries.
In the left panel, the gray region corresponds to the parameter space excluded by the Higgs invisible decay mentioned earlier.
In both panels, it is observed that for a fixed DM mass, a larger gauge coupling $g_\phi$ is required to reproduce the correct relic density as the reheating temperature $T_{\rm RH}$ decreases. This behaviour follows from the exponential suppression of the DM production rate at lower $T_{\rm RH}$, as indicated in eq.~\eqref{eq:thermal_sigma}. Consequently, a larger coupling is required to compensate for the suppressed production and yield the observed relic abundance.
In the same plane, we display $90\%$ confidence limit (C.L.) direct detection constraints from LZ \cite{LZ:2024zvo}, PandaX-4T \cite{PandaX-4T:2021bab} depicted by the light blue hatched region and magenta shaded region, respectively.
Due to the enhanced couplings in low $T_{\rm RH}$, we find a large portion of the parameter space is excluded.
We also showcase the potential future sensitivity from DARWIN \cite{DARWIN:2016hyl} and the neutrino floor represented by the blue dotted line and the orange dashed dot line, respectively.
Despite the stringent constraints, there still remains an allowed parameter space for such freeze-in DM that can be tested in future experiments like DARWIN.
Also, major parts of the relic satisfy the contours
lie below the neutrino floor.
\begin{figure}[!tbh]
 \centering
 \subfigure[]{\includegraphics[scale=0.4]{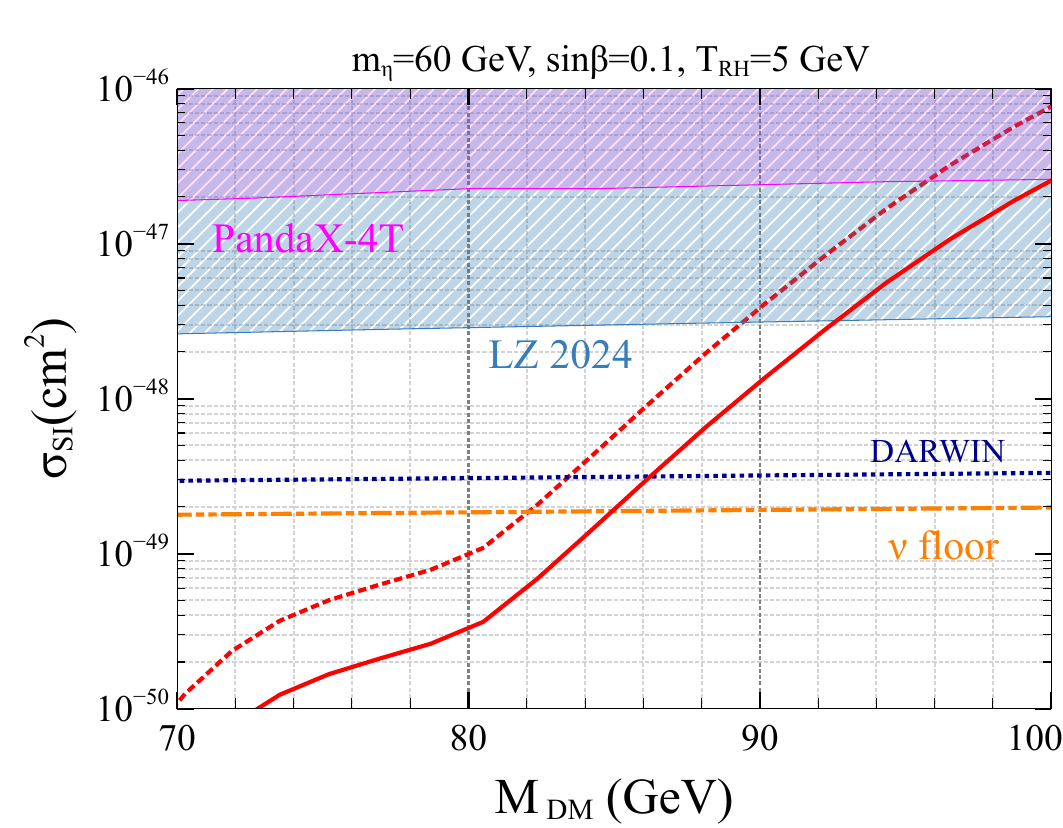}}
\subfigure[]{\includegraphics[scale=0.4]{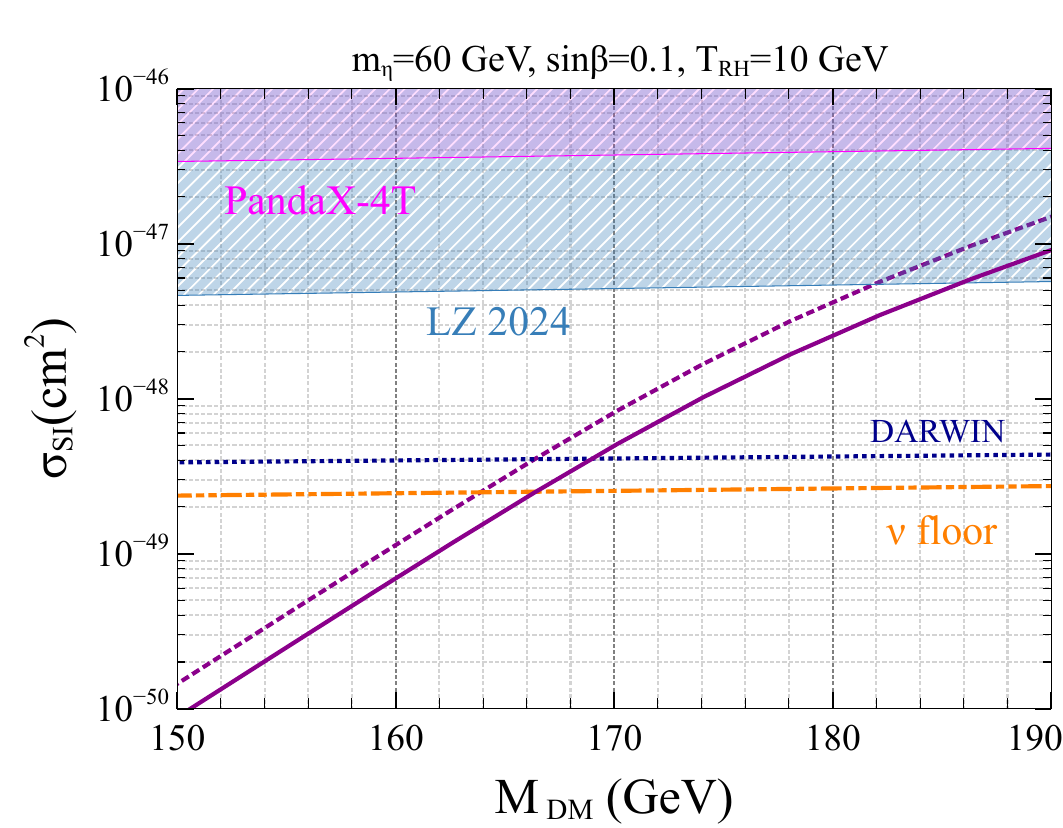}}
    \caption{ Direct search: comparison between the two DM cases for (a) $T_{\rm RH}=5$ GeV and (b) $T_{\rm RH}=10$ GeV.  Solid lines represent the parameter space for VDM in $SU(2)_{\rm HS}$ model and dotted lines represent the $U(1)_X$ model, respectively.}
    \label{fig:dd2}
\end{figure}

In Fig.\ref{fig:dd2} we highlight the importance of the 3 components of VDM particles realized in this $SU(2)_{\rm HS}$ model. 
In this plot We show the variation in DM-nucleon cross-section ($\sigma_{SI}$) with the DM mass for a fixed benchmark mediator masses, $m_\eta = 60~\text{GeV}$ and mixing angle $\sin\beta = 0.1$, considering two different values of reheating temperature $T_{\rm RH}=5$ GeV (left) and $T_{\rm RH}=10$ GeV (right) represented by red and magenta lines respectively.
In the same plane, we display the relic satisfying contour for the same set of benchmark parameters with only one component of DM (say $X_1$) represented by the dotted lines.
Note that, for a single component of DM (dotted line), a higher value of cross-section is required to satisfy the relic density for the same DM mass.
However, the direct detection rate depends only on the total DM density at the present day, independent of the number of components.
Hence, the direct detection constraints remain similar for both cases.
In such a scenario, the $SU(2)_{\rm HS}$ model can exhibit a larger allowed parameter space than that with a single DM component (e.g. $U(1)_X$ abelian VDM \cite{Khan:2025keb}).
For example, with $T_{\rm RH}=5$ GeV (10 GeV) DM mass up to $92$ GeV (184 GeV) is allowed from LZ 2024 in $SU(2)_{\rm HS}$ scenario, whereas for the other scenario, DM mass only up to $89$ GeV (180 GeV) is allowed. 
A similar conclusion can be drawn for the detection prospects in future experiments like DARWIN.

For the same set of benchmark parameters, we display the indirect detection constraints in Fig.\ref{fig:indirect_detection}.
Here also we fix $m_\eta = 60~\text{GeV}$ ({\bf left}) and $400~\text{GeV}$ ({\bf right}), and fixed mixing angle $\sin\beta = 0.1$.
Similar to the previous discussion,
in the left plot, the red, dark magenta, cyan, and brown solid lines signify $T_{\rm RH}=5$ GeV, 10 GeV, 20 GeV, and 40 GeV, respectively.
On the other hand, in the right plot, the red, dark magenta, cyan, and brown solid lines signify $T_{\rm RH}=30$ GeV, 60 GeV, 90 GeV, and 120 GeV, respectively.
For a fixed value of $g_\phi$, increasing the DM mass leads to a smaller annihilation cross section and hence a reduced relic abundance. As a result, a higher reheating temperature is required to reproduce the observed relic density. 
We have shown the  $90\%$ C.L. bounds on the thermal average cross-section for the process $X_iX_i\rightarrow W^+W^-$  from the Fermi-LAT \cite{Fermi-LAT:2015att}  and AMS \cite{Calore:2022stf} experiments, depicted by magenta and light blue shaded regions.
The Fermi-LAT \cite{Fermi-LAT:2015att} bounds are obtained from the $ \gamma$-ray observations of the dwarf spheroidal galaxies in the Milky Way.
AMS-02 \cite{Calore:2022stf} constraints are obtained by looking at the antiproton excess, respectively.
Note that the indirect signal of DM is roughly $\propto \langle \sigma v\rangle J$, while the $J$-factor scales as $\rho_{\rm DM}^2$ for a single component DM.
However, for our scenario with 3 degenerate DM components, the $J\propto \sum_{i=1,2,3} \rho_{X_i}^2$, with $\rho_{X_i}=\frac{1}{3}\rho_{\rm DM}$. 
Hence, to showcase the constraints for this scenario, we scale the limits obtained for single-component DM in existing literature by a multiplication factor of 3.
The top-most points for each colored line satisfying the observed limit 
correspond to the thermalization boundaries for each $T_{\rm RH}$, and we do not extend the lines above that.
For the chosen set of BP, the relic-satisfying contours lie below the existing indirect-detection constraints.
Indirect-detection constraints from other channels, such as $b\bar{b}$ and $\mu\bar{\mu}$, are also found to yield weaker limits on $\langle \sigma v\rangle$ compared to relic-satisfying contours. Hence, we do not show them for brevity. 

Another notable feature arises in this regime, where the reheating temperature is smaller than the DM mass, which significantly impacts the required gauge coupling. In the left panel of figure~\ref{fig:direct_detection}, the maximum allowed value of $g_\phi$ is $\mathcal{O}(10^{-2})$ for $M_{\rm DM} \sim 100~\text{GeV}$, increasing to $\mathcal{O}(10^{-1})$ for $M_{\rm DM} \sim 1~\text{TeV}$. In the right panel, $g_\phi$ can reach values as large as $\mathcal{O}(1)$ for $M_{\rm DM} \sim 3~\text{TeV}$ within the region allowed by direct detection constraints. Such coupling strengths are considerably larger than those typically encountered in conventional freeze-in scenarios. Consequently, these relatively large couplings are subject to stringent constraints from direct detection experiments, while simultaneously providing promising prospects for future detection.

\section{Conclusion}
\label{sec:concl}
 In this work, we consider a minimal beyond-the-Standard-Model (BSM) scenario 
where the SM particle spectrum 
is augmented by a hidden non-Abelian gauge symmetry
$SU(2)_{\rm HS}$, giving rise to three spin-1 gauge bosons 
$X_i (i = 1,2,3)$. The scalar sector includes an $SU(2)_{\rm HS}$
doublet scalar $\phi$ which is neutral under the SM gauge group 
but carries a charge under the hidden gauge symmetry. The spontaneous 
breaking of $SU(2)_{\rm HS}$ via the vacuum expectation value of
$\phi$ generates a common mass for the three vector bosons, 
while the resulting $SO(3)$ custodial symmetry ensures their stability,
allowing them to serve as vector dark matter candidates with the Higgs 
portal interactions with SM particles.    
The possibilities of DM at low $T_{\rm RH}$ have been investigated previously, including Higgs-portal
dark matter of different spins and a renormalizable Abelian vector-DM
realization at low reheating temperature~\cite{Cosme:2023xpa,Khan:2025keb,Arcadi:2024wwg}.
Likewise, the custodially stabilized hidden-$SU(2)$ vector-DM framework, with
three mass-degenerate gauge bosons and characteristic non-Abelian interactions,
is already well established  for freeze-out scenario~\cite{Hambye:2008bq,Baouche:2021wwa,Choi:2017zww}.
The new result of our analysis is therefore the quantitative realization
of the low-$T_{\rm RH}$ nonthermal regime specifically in this three-component
custodial $SU(2)_{\rm HS}$ framework. For this purpose, we consistently include
the ordinary Higgs-portal production channels, the mixed $h\eta$ initial state,
their reverse reactions, and the genuinely non-Abelian cross-species
semi-annihilation processes
$X_iX_j\leftrightarrow X_k h$ and
$X_iX_j\leftrightarrow X_k\eta$, and explicitly determine the thermalization
boundary beyond which the freeze-in description is no longer applicable.
We solve the relevant Boltzmann equations in the instantaneous reheating approximation with $Y_\chi(T_{\rm RH})=0$.
Although the semi-annihilation contribution is found to modify the final
abundance only mildly in the parameter region of interest, the threefold
custodial degeneracy has a direct quantitative consequence,
$Y_{\rm DM}=3Y_{X_1} $.
Thus, each degenerate vector species needs to account for only one third of
the observed total yield, so that, for otherwise comparable production
dynamics, the relic-density-compatible gauge coupling is smaller than in a
single-component vector-DM scenario (approximately by $1/\sqrt{3}$ when
$Y_X\propto g_\phi^2$). Together with the larger coupling required to
overcome the Boltzmann suppression for $T_{\rm RH}<m_X$, this multiplicity
effect leads to the quantitatively distinct relic-density and direct-detection
phenomenology obtained in our analysis.

Within this framework, we study the freeze-in production of VDM $X_i$ at reheating temperatures lower than DM mass ($T_{\rm RH}<M_{\rm DM}$).
In this limit, the DM phenomenology has a strong dependence on  $T_{\rm RH}$ in contrast to the usual freeze-in scenario since the production (SM SM $\to X_iX_i$) rate of $X_i$ suffers from a Boltzmann suppression ($\sim\exp{(-m_{\rm DM}/T_{\rm RH})}$).
Consequently, one needs to consider higher values of DM couplings than usual freeze-in DM to overcome the Boltzmann suppression and obtain the correct relic density.
This, in turn, helps realize a freeze in DM with comparatively stronger coupling that can be tested in current and future-generation DM search experiments, which is the key focus of this work.

We explicitly examine the dependence of DM abundance on various model parameters by numerically solving the Boltzmann equation, accounting for all relevant production channels.
We find that with higher gauge coupling ($g_\phi$) and the mixing angle ($\sin\beta$), the DM abundances increase as these parameters directly control the production cross-section.
More importantly, we notice that with an increase by a few factors in the ratio $m_{\rm DM}/T_{\rm RH}$, the DM abundances decrease by a few orders of magnitude due to the previously mentioned Boltzmann suppression. 
To compensate for this suppression, the correct relic density is obtained with a comparatively larger coupling than that used in the usual freeze-in scenario.
For example, for $m_{\rm DM}\sim 100$ GeV and $T_{\rm RH}=5$ GeV one requires $g_{\phi}=\mathcal{O}(10^{-1})$ in order to satisfy the observed relic density while fixing $m_\eta=60$ GeV and $\sin\beta=0.1$.
We also discuss the relevant theoretical and experimental constraints that restrict the scalar masses and the mixing angle.
Finally, we perform exhaustive numerical scans to obtain the contours that satisfy the observed relic density in the $m_{\rm DM}$ vs. $g_\phi$ plane for different values of $T_{\rm RH}$ and benchmark values of other model parameters.
We also discuss the direct detection constraints on the same coupling from experiments like LZ and PandaX-4T.
We find that the DM masses with the ratio $M_{DM}/T_{\rm RH}\sim 15-20$ are excluded from existing direct searches.
For completeness, we also discuss the indirect detection constraints, and we find that the most stringent bounds apply on the DM  annihilation channel to $W^+W^-$.
However, indirect detection constraints are found to be significantly weaker compared to the direct search limits.
Despite the stringent constraints, there still exists a viable parameter space that can accommodate a stable DM with the correct relic density.
Moreover, any new signal in future experiments can offer new insights in such a scenario, as the DM mass and the reheating temperature pose a strong correlation.
Thus, the freeze-in of VDM at low reheating temperatures in such $SU(2)_{\rm HS}$ setup 
may lead to significant insights into the experimental and phenomenological possibilities of DM model building.
In the future, extending VDM frameworks to larger hidden-sector gauge groups, such as $SU(N)_{\rm HS}$ with $N\geq3$ \cite{Frigerio:2022kyu}, could significantly broaden the paradigm and open up a rich landscape of novel dark matter phenomenology and model-building possibilities.

\section*{Acknowledgement}
SG acknowledges the support of a postdoctoral fellowship from IIT Kharagpur (Offer no. -  IIT/PDF/2025/PH/23).  X.-G.H was supported by the Fundamental Research Funds for the Central Universities, by the National Natural Science Foundation of the People’s Republic of China (Nos. 12090064, 12375088, and W2441004 ). 
SJ acknowledges the financial support from the National Natural Science Foundation of China (12425506 and 12375101) and the State Key Laboratory of Dark Matter Physics.

\appendix
\section{Parametrization}
\label{sec:apxA}
Here we present the parametrization of the quartic couplings in terms of the masses of the physical scalars. If $m_\eta>m_h$, then :
\begin{equation}
    m_{h,\eta}^{2}=\lambda_H\upsilon^{2}+\lambda_{\phi}\upsilon_{\phi}^{2}\mp(\lambda_{\phi}\upsilon_{\phi}^{2}-\lambda_H\upsilon^{2})/\cos{2\beta},
\end{equation}
and the quartic couplings can be expressed as
\begin{equation}
\lambda_H=\frac{m_{h}^{2}}{2\upsilon^{2}}\,\cos^2{\beta}+\frac{m_{\eta}^{2}}{2\upsilon^{2}}\,\sin^2{\beta},~~~~~\lambda_{\phi}=\frac{m_{h}^{2}}{2\upsilon_{\phi}^{2}}\,\sin^2{\beta}+\frac{m_{\eta}^{2}}{2\upsilon_{\phi}^{2}}\,\cos^2{\beta},~~~~~\lambda_{\phi H}=\frac{\sin{2\beta}}{2\upsilon\upsilon_{\phi}}\,\left(m_{\eta}^{2}-m_{h}^{2}\right).
\end{equation}
%
%

\section{Cross-sections for $h_i h_j\to X X$}
\label{sec:apxB}
The amplitude for the process $h_i h_j\to X X$ receives contributions from contact interaction, $s$-channel scalar exchange and, when allowed by the cubic $h_iXX$ vertices, $t$- and $u$-channel vector exchange:
\begin{equation}
\mathcal M_{ij}=\mathcal M_{ij}^{\rm cont}+\mathcal M_{ij}^{s}+\mathcal M_{ij}^{t}+\mathcal M_{ij}^{u}.
\end{equation}
The required vector $X$--scalar couplings are
\begin{equation}
g_{hX X}=-\frac{g_\phi^2v_\phi}{4}\sin{\beta},
\qquad
g_{\eta X X}=\frac{g_\phi^2v_\phi}{4}\cos{\beta},
\end{equation}
The total cross section can be written as,
\begin{equation}
    \begin{aligned}
        \sigma_{h_i h_j\rightarrow XX}&=\frac{1}{64\pi s}\sqrt{\frac{s(s-4m_X^2)}{(s-m_{h_i}^2-m_{h_j}^2)^2-4m_{h_i}^2m_{h_j}^2}}\left[\mathcal{C}_0+\mathcal{C}_1\frac{1}{\mathcal{A}^2-\mathcal{B}^2}+\mathcal{C}_2\frac{1}{\mathcal{B}}\mathrm{ln}\left(\frac{|\mathcal{A}+\mathcal{B}|}{|\mathcal{A}-\mathcal{B}|}\right)\right.\\
&\left.+\mathcal{C}_3(3\mathcal{A}^2+\mathcal{B}^2)+\mathcal{C}_4\mathcal{A}+\mathcal{C}_5\mathcal{A}(\mathcal{A}^2+\mathcal{B}^2)\right]
\label{eq:higgs_x}
    \end{aligned}
\end{equation}

For $h(k_1)h(k_2)\rightarrow X(p_1)X(p_2)$ the aforementioned terms will be,
\begin{equation}
		\begin{aligned}
			\mathcal{A}&=\frac{2m_h^2-s}{2m_X^2},\ \ \mathcal{B}=\frac{1}{2m_X^2}\sqrt{(s-4m_h^2)(s-4m_X^2)},\\
            C_s&= \frac{\cos{\beta}\rho_2v_\phi}{s-m_\eta^2+im_\eta \Gamma_\eta}-\frac{\sin{\beta}\rho_hv_\phi}{s-m_h^2+im_h \Gamma_h}+\sin^2{\beta}\\
        \end{aligned}
\end{equation}
    \begin{equation}
		\begin{aligned}
			\mathcal{C}_0&=2\left[\frac{|C_s|^2}{v_\phi^4}(12m_X^4-4sm_X^2+s^2)+\frac{4\sin^{4}\beta}{v_{\phi}^{4}}(-4m_{X}^{4}+4m_{X}^{2}m_{h}^{2}-6m_{X}^{2}s+m_{h}^{4}+2m_{h}^{2}s)\right.\\
			&\left.+\frac{8\sin^{2}\beta \Re(C_s)}{v_{\phi}^{4}}(-2m_{X}^{2}m_{h}^{2}+2m_{X}^{2}s-m_{h}^{2}s)\right]\\
        \end{aligned}
	\end{equation}
    \begin{equation}
		\begin{aligned}        
			\mathcal{C}_1&=\frac{4\sin^4\beta}{v_\phi^4m_X^4}(48m_X^8-32m_X^6m_h^2+24m_X^4m_h^4-16m_X^4m_h^2s+4m_X^4s^2-8m_X^2m_h^6+4m_X^2m_h^4s+m_h^8)\\
			\mathcal{C}_2&=4\left[\frac{\sin^4\beta}{m_X^2v_\phi^4(s-2m_h^2)}(-48m_X^8+32m_X^6m_h^2-8m_X^4m_h^4+4 m_X^4 s^2- 8 m_X^2 m_h^6+4 m_X^2 m_h^2 s^2\right.\\
			&\left.- 2 m_X^2 s^3+3 m_h^8-m_h^4s^2)+\frac{\sin^2\beta \Re(C_s)}{m_X^2v_\phi^4}(24m_X^6 + 2m_X^2(m_h^2-s)^2 + m_h^4 s - 4m_X^4(2m_h^2+s))\right]
\end{aligned}
	\end{equation}
    \begin{equation}
		\begin{aligned}      
			\mathcal{C}_3&=-\frac{1}{4}g_\phi^4\sin^4\beta\\
			\mathcal{C}_4&=\frac{8m_X^2\sin^2\beta}{v_\phi^4(2m_h^2-s)}\left[\sin^2\beta(8m_X^4-8m_X^2m_h^2+8m_X^2s+2m_h^4-4m_h^2s+s^2)\right.\\
			&\left.+\Re(C_s)(4m_X^2m_h^2-2m_X^2s+2m_h^2s-s^2)\right]\\
			\mathcal{C}_5&=\frac{8\sin^4\beta m_X^6}{v_\phi^4(2m_h^2-s)}\\
		\end{aligned}
	\end{equation}

For  $\eta(k_1)\eta(k_2)\rightarrow X(p_1)X(p_2)$ the terms in eq.\eqref{eq:higgs_x} will be
\begin{equation}
		\begin{aligned}
			\mathcal{A}&=\frac{2m_\eta^2-s}{2m_X^2},\ \ \mathcal{B}=\frac{1}{2m_X^2}\sqrt{(s-4m_\eta^2)(s-4m_X^2)},\\
            C_s&=\cos^2\beta+\frac{\rho_\eta v_\phi\sin\beta}{s-m_\eta^2+im_\eta \Gamma_\eta}-\frac{\rho_1 v_\phi \cos\beta}{s-m_h^2+im_h\Gamma_h}\\
            	\end{aligned}
	\end{equation}
    \begin{equation}
		\begin{aligned}
			\mathcal{C}_0&=2\left[\frac{|C_s|^2}{v_\phi^4}(12m_X^4-4sm_X^2+s^2)+\frac{4\cos^{4}\beta}{v_{\phi}^{4}}(-4m_X^{4}+4m_{X}^{2}m_\eta^{2}-6m_{X}^{2}s+m_\eta^{4}+2m_\eta^{2}s)\right.\\
			&\left.+\frac{8\cos^{2}\beta \Re(C_s)}{v_{\phi}^{4}}(-2m_{X}^{2}m_\eta^{2}+2m_{X}^{2}s-m_\eta^{2}s)\right]\\
            \mathcal{C}_1&=\frac{4\cos^4\beta}{v_\phi^4m_X^4}(48m_X^8-32m_X^6m_\eta^2+24m_X^4m_\eta^4-16m_X^4m_\eta^2s+4m_X^4s^2-8m_X^2m_\eta^6+4m_X^2m_\eta^4s+m_\eta^8)\\
        \end{aligned}
	\end{equation}
    \begin{equation}
		\begin{aligned}        
			\mathcal{C}_2&=4\left[\frac{\cos^4\beta}{m_X^2v_\phi^4(s-2m_\eta^2)}(-48m_X^8+32m_X^6m_\eta^2-8m_X^4m_\eta^4+4 m_X^4 s^2- 8 m_X^2 m_\eta^6+4 m_X^2 m_\eta^2 s^2\right.\\
			&\left.- 2 m_X^2 s^3+3 m_\eta^8-m_\eta^4s^2)+\frac{\cos^2\beta \Re(C_s)}{m_X^2v_\phi^4}(24m_X^6 + 2m_X^2(m_\eta^2-s)^2 + m_\eta^4 s - 4m_X^4(2m_\eta^2+s))\right]\\
            \mathcal{C}_3&=-\frac{1}{4}g_\phi^4\cos^4\beta\\
			\mathcal{C}_4&=\frac{8m_X^2\cos^2\beta}{v_\phi^4(2m_\eta^2-s)}\left[\cos^2\beta(8m_X^4-8m_X^2m_\eta^2+8m_X^2s+2m_\eta^4-4m_\eta^2s+s^2)\right.\\
			&\left.+\Re(C_s)(4m_X^2m_\eta^2-2m_X^2s+2m_\eta^2s-s^2)\right]\\
			\mathcal{C}_5&=\frac{8\cos^4\beta m_X^6}{v_\phi^4(2m_\eta^2-s)}\\
\end{aligned}
	\end{equation}

On the other hand, for the process $h(k_1)\eta(k_2)\rightarrow X(p_1)X(p_2)$ the terms in eq.\eqref{eq:higgs_x} will be,
	\begin{equation}
		\begin{aligned}
			\mathcal{A}&=\frac{m_h^2+m_\eta^2-s}{2m_X^2},\ \ \mathcal{B}=\frac{1}{2m_X^2}\sqrt{1-\frac{4m_X^2}{s}}\sqrt{(s-m_h^2-m_\eta^2)^2-4m_h^2m_\eta^2},
            \end{aligned}
	\end{equation}
    \begin{equation}
		\begin{aligned} 
            C_s&=(-\sin\beta\cos\beta+\frac{\rho_1\cos\beta v_\phi}{s-m_\eta^2+im_\eta \Gamma_\eta}-\frac{\rho_2\sin\beta v_\phi}{s-m_h^2+im_h \Gamma_h})\\
			\mathcal{C}_0&=2\frac{|C_s|^2}{v_\phi^4}(s^2-4m_X^2s+12m_X^4)\\
			&+\frac{\sin^2 2\beta}{v_\phi^4}(-8 m_X^4 + 4 m_X^2 m_\eta^2 + m_\eta^4 + 4 m_X^2 m_h^2 + m_h^4 - 12 m_X^2 s + 
			2 m_\eta^2 s + 2 m_h^2 s)\\
			&+4\Re(C_s)\frac{\sin2\beta}{v_\phi^4}(2 m_X^2 m_\eta^2 + 2 m_X^2 m_h^2 - 4 m_X^2 s + m_\eta^2 s + m_h^2 s)
            \end{aligned}
	\end{equation}
    \begin{equation}
		\begin{aligned} 
			\mathcal{C}_1&=\frac{\sin^2 2\beta}{m_X^{4}\upsilon_{\phi}^{4}}(48m_X^{8}-16m_X^{6}m_{\eta}^{2}-16m_X^{6}m_h^{2}+8m_X^{4}m_{\eta}^{4}+8m_X^{4}m_{\eta}^{2}m_h^{2}-8m_X^{4}m_{\eta}^{2}s+8m_X^{4}m_h^{4}\nonumber\\
			& -8m_X^{4}m_h^{2}s 
			+4m_X^{4}s^{2}-4m_X^{2}m_{\eta}^{4}m_{h}^{2}-4m_X^{2}m_{\eta}^{2}m_{h}^{4}+4m_X^{2}m_{\eta}^{2}m_{h}^{2}s+m_{\eta}^{4}m_{h}^{4}),\nonumber \\
            \end{aligned}
	\end{equation}
    \begin{equation}
		\begin{aligned} 
			\mathcal{C}_2& =\frac{2\Re(C_s) \sin 2\beta}{m_X^{2}\upsilon_{\phi}^{4}}(-24m_X^{6}+4m_X^{4}m_{\eta}^{2}+4m_X^{4}m_{h}^{2}+4m_X^{4}s-2m_X^{2}m_{\eta}^{4}+2m_X^{2}m_{\eta}^{2}m_{h}^{2}+2m_X^{2}m_{\eta}^{2}s-2m_X^{2}m_{h}^{4}\nonumber \\
			& +2m_X^{2}m_{h}^{2}s-2m_X^{2}s^{2}-m_{\eta}^{2}m_{h}^{2}s),\nonumber \\
			& + \frac{\sin^2 2\beta}{m_X^{2}\upsilon_{\phi}^{4}(m_h^2+m_\eta^2-s)}(48 m_X^8 - 16 m_X^6 m_\eta^2 + 8 m_X^4 m_\eta^4 - 2 m_X^2 m_\eta^6\\
			& - 
			16 m_X^6 m_h^2 - 8 m_X^4 m_\eta^2 m_h^2 + 6 m_X^2 m_\eta^4 m_h^2 - 
			m_\eta^6 m_h^2 + 8 m_X^4 m_h^4 + 6 m_X^2 m_\eta^2 m_h^4 \\
			&- m_\eta^4 m_h^4 - 
			2 m_X^2 m_h^6 - m_\eta^2 m_h^6 + 2 m_X^2 m_\eta^4 s - 4 m_X^2 m_\eta^2 m_h^2 s + 
			2 m_X^2 m_h^4 s \\
			&- 4 m_X^4 s^2 - 2 m_X^2 m_\eta^2 s^2 - 2 m_X^2 m_h^2 s^2 + 
			m_\eta^2 m_h^2 s^2 + 2 m_X^2 s^3 )\\
            \mathcal{C}_3&=-\frac{1}{4}g_\phi^4\cos^2\beta\sin^2\beta\\
            \end{aligned}
	\end{equation}
    \begin{equation}
		\begin{aligned} 
			\mathcal{C}_4&=-\frac{4\sin 2\beta \Re(C_s)m_X^2}{v_\phi^4}(2m_X^2+s)-\frac{2\sin^2 2\beta m_X^2}{\upsilon_\phi^4}(m_\eta^2+m_h^2)+\frac{2\sin^2 2\beta m_X^2}{\upsilon_\phi^4(m_h^2+m_\eta^2-s)}\\
			&(8 m_X^4 - 4 m_X^2 m_h^2 + m_h^4 - 4 m_X^2 m_\eta^2 + 
			4 m_h^2 m_\eta^2 + m_\eta^4 + 8 m_X^2 s - 3 m_h^2 s - 
			3 m_\eta^2 s + s^2)\\
			\mathcal{C}_5&=\frac{2m_X^6\sin^2 2\beta}{(m_h^2+m_\eta^2-s)v_\phi^4}\\
		\end{aligned}
	\end{equation}
\section{Interaction Lagrangian of scalar sector}
\label{sec:apxC}
In the mass eigenstate basis, the interaction terms in the Lagrangian is :
\begin{eqnarray}
    \mathcal{L}\supset -\frac{1}{6}\rho_h h^3 - \frac{1}{6}\rho_\eta \eta^3 - \frac{1}{2}\rho_1\eta^2h - \frac{1}{2}\rho_2h^2\eta
\end{eqnarray}
where the scalar trilinear couplings are given by 
\begin{align}
\rho_1 &= -6 \lambda_\phi v_\phi \cos^2\beta \sin\beta 
+ 6 \lambda_H v \sin^2\beta \cos\beta \notag\\
&\quad + \lambda_{\phi H} \left( v \cos^3\beta - v_\phi \sin^3\beta 
- 2 v \cos\beta \sin^2\beta + 2 v_\phi \cos^2\beta \sin\beta \right), \\[6pt]
\rho_2 &= 6 \lambda_\phi v_\phi \sin^2\beta \cos\beta 
+ 6 \lambda_H v \cos^2\beta \sin\beta \notag\\
&\quad + \lambda_{\phi H} \left( v \sin^3\beta + v_\phi \cos^3\beta 
- 2 v \cos^2\beta \sin\beta - 2 v_\phi \sin^2\beta \cos\beta \right), \\[6pt]
\rho_\eta &= 6 \lambda_\phi v_\phi \cos^3\beta 
+ 6 \lambda_H v \sin^3\beta 
+ 3 \lambda_{\phi H} \cos\beta \sin\beta \left( v \cos\beta + v_\phi \sin\beta \right), \\[6pt]
\rho_h &= -6 \lambda_\phi v_\phi \sin^3\beta 
+ 6 \lambda_H v \cos^3\beta 
+ 3 \lambda_{\phi H} \cos\beta \sin\beta \left( v \sin\beta - v_\phi \cos\beta \right).
\end{align}

\section{Semi-annihilation cross-section and contribution in collision term}
\label{sec:apxD}

\begin{figure}[!tbh]
    \centering
    \includegraphics[scale=0.5]{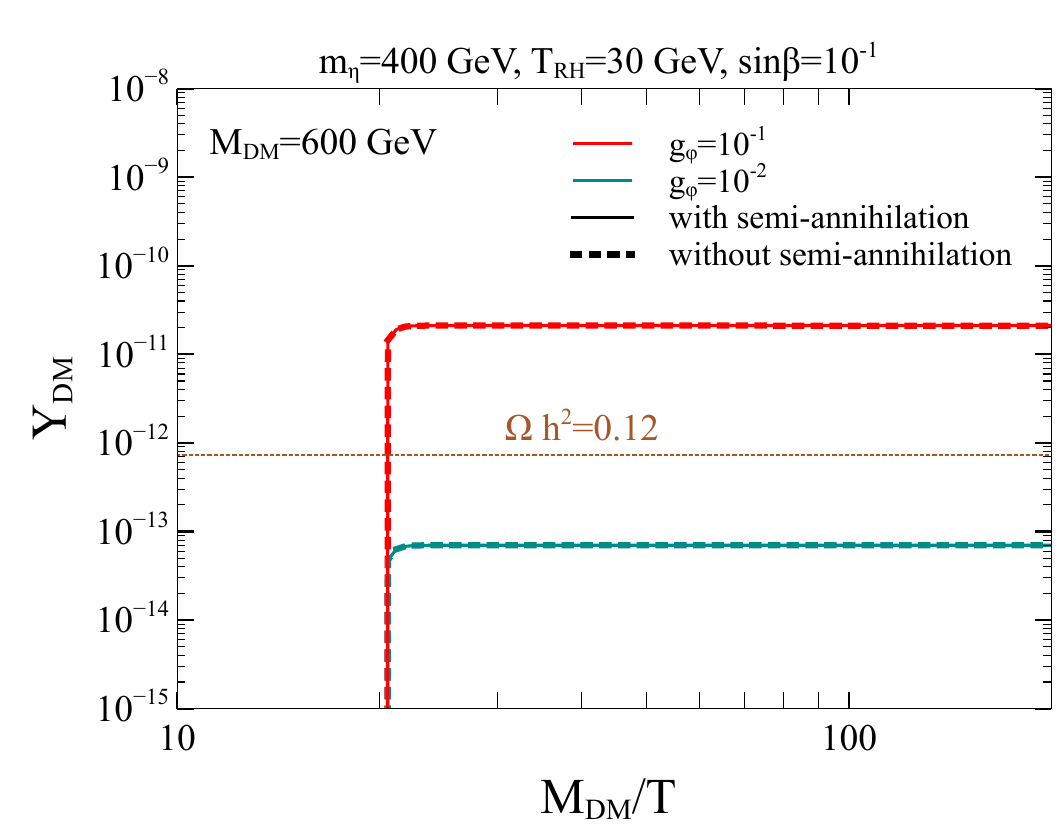}
    \caption{Evolution of co-moving abundance with (solid) and without (dashed) the contribution from semi-annihilation $X_iX_j\to\eta X_k(h X_k)$ for different values of $g_\phi=10^{-1}$ (red) and $10^{-2}$ (cyan).}
    \label{fig:semi}
\end{figure}

The cross sections for the semi-annihilation can be written as
\begin{equation}
    \begin{aligned}
    \sigma_{X_1 X_2\rightarrow h_iX_3}&=\frac{1}{32\pi s}\sqrt{\frac{(s-m_X^2-m_{h_i}^2)^2-4m_X^2m_{h_i}^2}{s(s-4m_X^2)}}\left[\mathcal{C}_0+\mathcal{C}_1\frac{1}{\mathcal{A}^2-\mathcal{B}^2}+\mathcal{C}_2\frac{1}{\mathcal{B}}\mathrm{ln}\left(\frac{|\mathcal{A}+\mathcal{B}|}{|\mathcal{A}-\mathcal{B}|}\right)\right.\\
&\left.+\mathcal{C}_3(3\mathcal{A}^2+\mathcal{B}^2)+\mathcal{C}_4\mathcal{A})\right].
\label{eq:semi}
    \end{aligned}
\end{equation}
The parameters are given by
\begin{equation}
	\begin{aligned}
		\mathcal{A}&=\frac{m_X^2+m_{h_i}^2-s}{2m_X^2},\ \ \mathcal{B}=\frac{1}{2m_X^2}\sqrt{1-\frac{4m_X^2}{s}}\sqrt{(s-m_X^2-m_{h_i}^2)^2-4m_X^2m_{h_i}^2},\\
    \end{aligned}
\end{equation}
\begin{equation}
    \begin{aligned}
    \mathcal{C}_0 & = \frac{2m_X^2 q^2}{9(m_X^2-s)^2 v_\phi^4}(9 m_X^2 s^2 - 8 m_{h_i}^4 s + 38 m_X^2 m_{h_i}^2 s - 414 m_A^4 s - 7 m_X^2 m_{h_i}^4 + 58 m_X^4 m_{h_i}^2 + 9 m_X^6) \\
    \mathcal{C}_1&=\frac{4 q^2 }{3v_\phi^4}(12s^2-12m_{h_i}^2s-24m_X^2s-5m_{h_i}^4-44m_{h_i}^2m_X^2-120m_X^4),\\
    \mathcal{C}_2&=\frac{4 q^2}{9 v_\phi^2 (m_X^2-s)(m_{h_i}^2+m_X^2-s)} \left(18 m_X^2 s^3- 252 m_X^4 s^2 + m_X^2 m_{h_i}^2 s^2 - 4 m_{h_i}^4 s^2 \right.\\
    &\left.+ 450 m_X^6 s + 196 m_X^4 m_{h_i}^2 s - 11 m_X^2 m_{h_i}^4 s + 4 m_{h_i}^6 s -216 m_X^8 + 199 m_X^6 m_{h_i}^2 \right.\\
    &\left.- 141 m_X^4 m_{h_i}^4 + 7m_X^2 m_{h_i}^6 - 2 m_{h_i}^8 \right),\\
    \mathcal{C}_3 &= \frac{8 m_X^8 q^2}{3(m_X^2 - s)^2 v_\phi^4},\\
    \mathcal{C}_4 & = \frac{8 m_X^6 (s-m_X^2-m_{h_i}^2)q^2}{(m_X^2-s)^2 v_\phi^4},
    \end{aligned}
\end{equation}
with $q=\cos\beta$ for $X\eta$ final state and $q=\sin\beta$ for $X h$ final state.

The thermal average cross-section can be obtained from eq.\eqref{eq:thermal_sigma} and subsequently we substitute it into eq.\eqref{eq:boltz} to obtain the co-moving abundance.
In Fig.\ref{fig:semi} we showcase the evolution of co-moving abundances with (solid) and without (dashed) the contribution from semi-annihilation $X_iX_j\to\eta X_k(hX_k)$, taking $m_\eta=400$ GeV,  $\sin\beta=10^{-1}$, $T_{\rm RH}=30$ GeV and $M_{\rm DM}=600$ GeV.
Abundances for different values of $g_\phi=10^{-1}$ (red) and $10^{-2}$ (cyan) are shown by different colored lines.
We observe that, even with modest values of $g_\phi$, neglecting the semi-annihilation does not affect the evolution much.
Nevertheless, for consistency, we implement the semi-annihilation processes throughout our numerical analysis.
\bibliography{ref}
\bibliographystyle{JHEP}

\end{document}